\documentclass[twocolumn]{aastex701}

\usepackage{subcaption}
\usepackage{multirow}
\usepackage{makecell}
\usepackage{threeparttable}
\usepackage[T1]{fontenc}

\begin{document}

\title{A First Look at Astrometric Outer Companions to Super-Earths and Sub-Neptunes
\newline using Gaia DR2 and DR3} 



\newcommand{\nrc}{NRC Herzberg Astronomy and Astrophysics,
5071 West Saanich Road,
Victoria, BC, V9E 2E7, Canada}
\newcommand{\ucsd}{Department of Astronomy \& Astrophysics,  University of California, San Diego, La Jolla, CA 92093, USA}
\newcommand{\nmsu}{Department of Astronomy, New Mexico State University, 1320 Frenger Mall, Las Cruces, NM 88003, USA}
\newcommand{\uvic}{Department of Physics and Astronomy, University of Victoria,
3800 Finnerty Road, Elliott Building,
Victoria, BC, V8P 5C2, Canada}
\newcommand{\mac}{Department of Physics \& Astronomy, McMaster University, 1280 Main Street West, Hamilton, ON, L8S 4L8, Canada}
\newcommand{\caltech}{Department of Astronomy, California Institute of Technology, Pasadena, CA 91125, USA}
\newcommand{\hawaii}{Institute for Astronomy, University of Hawaii at M\=anoa, 2680 Woodlawn Drive, Honolulu, HI 96822, USA}
\newcommand{\utaustin}{Department of Astronomy, The University of Texas at Austin, Austin, TX 78712, USA}
\newcommand{\ucsc}{Department of Astronomy \& Astrophysics, University of California, Santa Cruz, CA 95064, USA}
\newcommand{\ucsb}{Department of Physics, University of California, Santa Barbara, CA 93106, USA}
\newcommand{\ucla}{Department of Earth, Planetary, and Space Sciences, University of California, Los Angeles, CA 90095, USA}

\author[0009-0008-1229-3230]{Alexandra Rochon}
\affiliation{Department of Physics \& Astronomy, McMaster University, 1280 Main St W, Hamilton, ON, L8S 4L8, Canada}
\email[show]{rochoa3@mcmaster.ca}

\author[0000-0002-1845-2617]{Emma V. Turtelboom}
\altaffiliation{William \& Caroline Herschel Fellow}
\affiliation{Department of Physics \& Astronomy, McMaster University, 1280 Main St W, Hamilton, ON, L8S 4L8, Canada}
\email{turtlebe@mcmaster.ca}

\author[0000-0001-5383-9393]{Ryan Cloutier}
\affiliation{Department of Physics \& Astronomy, McMaster University, 1280 Main St W, Hamilton, ON, L8S 4L8, Canada}
\email{ryan.cloutier@mcmaster.ca}

\author[0000-0001-5684-4593]{William Thompson}
\affiliation{\nrc}
\email{william.thompson@nrc-cnrc.gc.ca}

\author[0000-0001-7402-8506]{Alexandre Bouchard-Côté}
\affiliation{Department of Statistics, University of British Columbia, Vancouver, BC V6T 1Z4, Canada}
\email{bouchard@stat.ubc.ca}

\author[0000-0003-1499-0191]{Trevor Campbell}
\affiliation{Department of Statistics, University of British Columbia, Vancouver, BC V6T 1Z4, Canada}
\email{trevor@stat.ubc.ca}

\author[0000-0002-6773-459X]{Doug Johnstone}
\affiliation{\nrc}
\affiliation{\uvic}
\email{Douglas.Johnstone@nrc-cnrc.gc.ca}

\author[0000-0003-3430-3889]{Jessica Speedie}
\altaffiliation{Heising-Simons Foundation 51 Pegasi b Fellow}
\affiliation{Department of Earth, Atmospheric, and Planetary Sciences, Massachusetts Institute of Technology, Cambridge, MA 02139, USA}
\email{jspeedie@mit.edu}

\author[0000-0002-4290-6826]{Judah Van Zandt}
\affiliation{\ucsb}
\email{judahvz@ucsb.edu}

\author[0000-0001-9582-4261]{Dori Blakely}
\affiliation{\uvic}
\affiliation{\nrc}
\email{blakelyd@uvic.ca}

\author[0000-0002-9820-1884]{Simon Blouin}
\affiliation{\uvic}
\email{sblouin@uvic.ca}

\author[0000-0003-2649-2288]{Brendan P. Bowler}
\affiliation{\ucsb}
\email{bpbowler@ucsb.edu}

\author[0000-0001-5173-2947]{Clarissa R. Do \'O}
\affiliation{\caltech}
\email{cdoo@caltech.edu}

\author[0009-0007-6766-2040]{Andre Fogal}
\affiliation{\uvic}
\email{afogal@uvic.ca}

\author[0000-0003-4557-414X]{Kyle Franson}
\altaffiliation{NHFP Sagan Fellow}
\affiliation{\ucsc}
\email{kfranson@ucsc.edu}

\author[0009-0009-2223-2404]{Kaitlyn Hessel}
\affiliation{\uvic}
\email{khessel@uvic.ca}

\author[0000-0002-4164-4182]{Christian Marois}
\affiliation{\nrc}
\affiliation{\uvic}
\email{Christian.Marois@nrc-cnrc.gc.ca}

\author[0000-0001-6975-9056]{Eric Nielsen}
\affiliation{\nmsu}
\email{nielsen@nmsu.edu}

\author[0009-0008-9687-1877]{William Roberson}
\affiliation{\nmsu}
\email{wcr@nmsu.edu}

\author[0000-0003-2233-4821]{Jean-Baptiste Ruffio}
\affiliation{\ucsd}
\email{jruffio@ucsd.edu}

\author[0000-0002-6618-1137]{Jerry W. Xuan}
\altaffiliation{Heising-Simons Foundation 51 Pegasi b Fellow}
\affiliation{\ucsb}
\email{jxuan@ucsb.edu}

\author[0000-0002-2696-2406]{Jingwen Zhang}
\affiliation{\ucsb}
\email{jwzhang@ucsb.edu}


\begin{abstract}
Massive outer companions (OCs) are believed to promote, inhibit, or otherwise regulate the formation and composition of inner small planets (ISPs) both in the solar system and in exoplanetary systems.
To assess how, we leverage calibrated absolute astrometry from Gaia DR2-DR3 of 7,321 FGKM dwarfs and present measurements of the relative occurrence rate of OCs in systems hosting known ISPs (i.e. P(OC$|$ISP)/P(OC)), including systems with inner super-Earths only, inner sub-Neptunes only, and across a range of spectral types and stellar metallicities.
We search for OCs in pairs of planet and reference stellar samples using the orbital fitting framework \texttt{Octofitter} and characterize the occurrence of OCs in mass--semi-major axis space under a Bayesian hierarchical model. We find an excess of OCs in systems with ISPs ($1.77\pm0.49$), independently supporting results from radial velocity surveys for the first time using astrometry. 
We recover similar excesses in both super-Earth (SE; $2.02\pm0.52$) and sub-Neptune systems (SN; $1.97\pm0.50$), highlighting that OCs do not inhibit the formation of ISPs, regardless of planet type. We argue that the consistency across SE and SN systems supports ex-situ formation, wherein ISPs form rapidly at larger separations prior to OCs opening disk gaps, thus providing evidence of Type I migration of SEs and SNs.
We also evaluate how these results vary across stellar mass and metallicity, finding that M dwarfs exhibit an OC excess that is $1.41\pm0.78\times$ stronger than that around FGK dwarfs, and not finding a significant dependence on metallicity, contrary to previous radial velocity surveys.

\end{abstract}



\section{Introduction} \label{sec:Intro}

Massive planets and substellar companions shape planetary system architectures. In the solar system, the gravitational influences of Jupiter and Saturn played dominant roles in the formation of the inner terrestrial planets \citep{Tsiganis2005, walsh2011,Morbidelli2012} and may have driven the migration of water-rich planetesimals onto inner orbits, delivering water to the early Earth \citep{obrien2006, Raymond2006,  raymond_izodoro2017}. Beyond the solar system, the conditional occurrence rate of cold gas giants (GGs) given inner small planets (ISPs) provides a key window into understanding how massive outer companions (OCs) shape the formation of the super-Earths and sub-Neptunes, the most common planets in the Milky Way \citep[e.g.][]{Fressin2013,Dressing2015,Hsu2018,Hsu2020,Kunimoto2020,Gillis2026}. 

The expected relationship between OCs and ISPs based on formation theories therefore remains cloudy. 
GGs can carve gaps in the protoplanetary disk, forming pressure bumps that suppress the inward flux of pebbles from the outer disk \citep{bitsch2021}. These pebble traps lead to pile-ups of dust exterior to the gap, which are observable as gap and ring structures in sub-mm continuum images of many disks, except the most compact \citep[e.g.][]{Fedele2017,Bosschaart2026,Curone2026}. Several models have shown that ISP formation strongly depends on the availability of pebbles drifting inwards from the outer disk \citep[e.g.][]{Lambrechts2019,bitsch2021, Mulders2021,Chachan2022,2025Danti,Sandhaus2025}. Therefore, OCs may both inhibit inner ISP formation by suppressing pebble delivery \citep{bitsch2021} and prevent the inward migration of ISPs that formed exterior to the OC \citep{Izidoro2015}, consequently predicting an anti-correlation between OC and ISP occurrence. Conversely, pressure bumps are known to be imperfect, leaky traps with a non-zero fraction of solids able to cross the gap and enter the inner system \citep{Huang2025, Sandhaus2025, Houge2026}. Indeed, disk solid masses that are sufficient to form OCs of hundreds of Earth masses also have sufficient material in the inner disk to form ISPs that are at least an order of magnitude less massive \citep{Chachan2022}. 

Previous radial velocity (RV) studies have worked to establish the correlation between ISPs and cold GGs \citep[e.g.][and references below]{Bryan2016,ZhuWu2018,Bryan2019,Bonomo2023,Bryan2024,Bonomo2025,VanZandt2025,Quanyi2026,Ruggieri2026}. The majority of these analyses have found evidence of enhanced GG occurrence in systems hosting ISPs relative to field stars, implying that the formation of GGs and ISPs are not mutually exclusive. Follow up work has suggested that this observed correlation is most evident around metal-rich, massive stars \citep{Zhu2024,Bryan2025,DiChang2026}. The reproducibility of these results by multiple groups suggests that the reported correlations are robust, despite the RV survey designs being necessarily restricted to relatively small sample sizes not exceeding more than 217 ISP systems. Consequently, most RV studies consider all ISPs as a single population without distinguishing between super-Earths (SEs) and sub-Neptunes (SNs). Recent exceptions include \cite{DiChang2026}, whose ISP sample was limited to 28 SEs plus 52 SNs, and \cite{Tang2026}, who studied 43 ISP systems with mass and radius measurements, the majority of which would be considered volatile-rich SNs. These studies compiled cold GG detections from the literature and infered that SNs show a greater enhancement in OC occurrence compared to SEs, although these inferences are limited by small sample sizes and the inhomogeneous nature of their GG planet searches.


In this study, we make use of the large catalogue of astrometric measurements from the Gaia mission \citep{GaiaCollaboration2016} to greatly expand upon the limited sample sizes of RV studies. We conduct a systematic search for OCs in calibrated, absolute Gaia DR2 and DR3 astrometry from the G23H catalog \citep{Thompson2026} to investigate the relative occurrence of OCs and ISPs. With our statistics, we are able to probe the difference between the SE and SN populations separately and across a range of stellar masses and metallicities. Our work provides a first astrometric look at the correlations between OCs and various ISPs in advance of Gaia's fourth data release on December 2, 2026, which is expected to yield thousands of exoplanet discoveries \citep{Lammers2026}. 

Our paper is structured as follows. We describe our methodology, including our target sample definitions, our OC search procedure, and our calculation of OC occurrence rates using a hierarchical Bayesian model in Section~\ref{sec:methods}. We present our results in Section~\ref{sec:results} and discuss the implications of our findings in Section~\ref{sec:Discussion}. We present our conclusions in Section~\ref{sec:conclusion}.

\section{Methods} \label{sec:methods}
 In this study, we aim to quantify the impact of OCs on ISPs by measuring and comparing the conditional occurrence rate of OCs in the presence of known inner super-Earths (SEs) and/or sub-Neptunes (SNs), and around stars of varying spectral type and metallicity. We define our OCs of interest as objects with masses between $1-80\, M_{\rm{Jup}}$ and semi-major axes between $1-20$ AU. This mass range captures Jovian-mass planets and extends up to roughly the hydrogen-burning limit. The 1 AU lower limit is commonly adopted in previous studies to define cold, or outer companions, while also avoiding overlap with the known ISPs and dynamical interactions between ISPs and the closest in OCs. We select an upper limit of 20 AU based on preliminary tests that indicated that the detection sensitivity of our data drops severely beyond this limit.

We leverage a large stellar sample with available astrometry, allowing us to investigate meaningful trends for SEs and SNs separately. We do not report absolute occurrence rates, i.e. sensitivity-corrected measurements, but instead present \textit{relative} occurrence rates that are independent of detection sensitivity. In other words, we measure the ratio of OC occurrence in systems hosting known ISPs to the occurrence of OCs around field stars: 

\begin{equation}
    \frac{\textrm{P(OC}|\textrm{ISP)}}{\textrm{P(OC)}}.
    \label{eq:rel}
\end{equation}

\noindent In the case that our detection sensitivity to OCs in both terms is identical, then we can robustly measure the \emph{relative} OC occurrence rate without having to quantify detection sensitivity.
To ensure the detection sensitivities across both our planet (numerator) and reference (denominator) samples are equivalent, we match the samples' distributions of relevant parameters that are known to affect astrometric sensitivity to OCs. We describe how we construct our planet and reference samples in Sections~\ref{sect:planet_sample} and \ref{sec:ref-sample}, respectively.

\subsection{Planet Sample Selection}\label{sect:planet_sample}

We construct samples of stars that host known ISPs, focusing separately on transiting SE and SN hosts.
In this study, we adopt convenient definitions of SEs and SNs based solely on their radii of $R_p < 1.6\, R_{\oplus}$ and $1.6 \le R_p/R_{\oplus} < 3$, respectively. These definitions are chosen for simplicity given that precise classifications of SEs and SNs relies on knowledge of their bulk compositions, which would require systematic mass characterization that does not exist for most planets of interest. Our definitions broadly hold given the existence of the radius valley between $\approx 1.6-1.9 \, R_\oplus$ around FGK and early M stars \citep[e.g.][]{Rogers2015,Fulton2017,Wu2019,cloutier_menou2020,hardegreeullman_2020,LuquePalle2022,Petigura2022}. We do not impose a lower radius limit on SE planets such that this planet sample also contains canonical Earths and sub-Earths, although we refer to these terrestrial objects collectively as SEs for convenience. 

We begin by selecting TESS Objects of Interest (TOIs) according to the radius cuts above and with orbital periods $<30$ days to restrict our analysis to \textit{inner} small planets. We retain all planet hosts with stellar masses $M_\star \le 1.5\, M_{\odot}$, surface gravities $\log{g} > 3.5$, and distances $d < 300$ pc, corresponding to nearby main sequence stars with spectral types later than $\sim$ F0V. We then cut on TOI dispositions, retaining only known planets (KP), confirmed planets (CP), and planet candidates (PC), which we colloquially refer to as planets. 
While the TOIs in our sample have been vetted \citep{Guerrero2021}, although we cannot guarantee that they have zero false positive and false alarm rates. Deriving a bias-corrected planet sample from the full TOI list is beyond the scope of this work and remains a caveat of our results. We adopt stellar parameters from the TESS Input Catalog \citep[TIC-8;][]{Stassun2019} and use them to derive corresponding planetary parameters from direct transit observables provided on ExoFOP \citep{Christiansen2025}.

\begin{figure}
    \centering
    \includegraphics[width=\hsize]{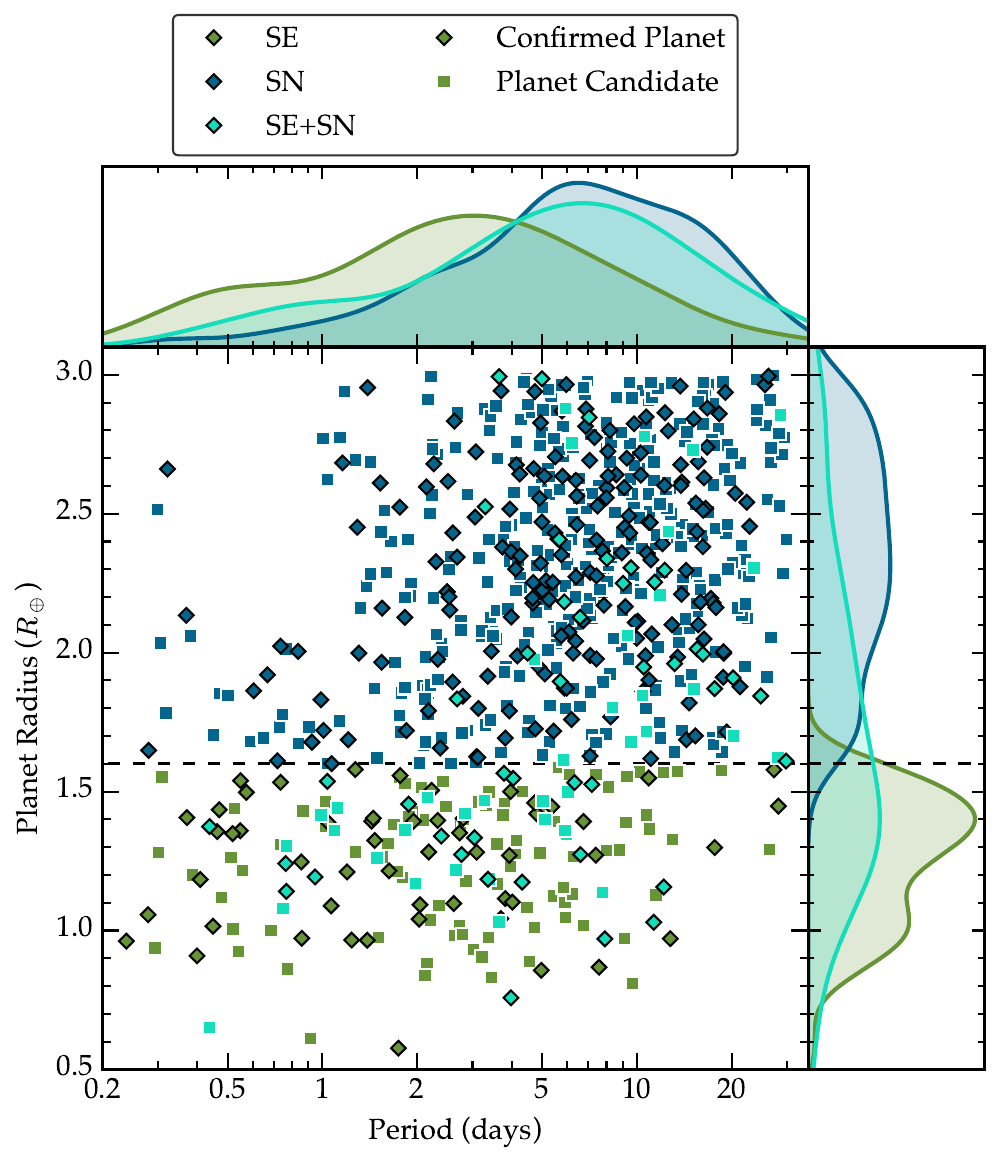}
    \caption{Planet radius vs. orbital period for the planets and planet candidates in our super-Earth-only sample (SE, green), sub-Neptune-only sample (SN, dark blue), and the super-Earth plus sub-Neptune sample (SE+SN, light blue), included in our ALL ISP sample (see text). Planet parameters are sourced from ExoFOP. Confirmed planets are shown as diamonds with black edges, while candidates are shown as squares without edge colours. The horizontal dashed line at $1.6\, R_\oplus$ marks our canonical distinction between SEs and SNs. Kernel density estimates of each sample in period and radius are shown on the top and right of the main plot.}
    \label{fig:planet}
\end{figure}

We construct three planet samples and depict their joint radius-period distribution in Figure~\ref{fig:planet}.

\begin{enumerate}
    \item \emph{SE}: all known transiting planets in the system satisfy our SE radius criterion ($R_p < 1.6\, R_{\oplus}$). This sample contains 134 systems.
    \item \emph{SN}: all known transiting planets in the system satisfy our SN radius criterion ($1.6 \le R_p/R_{\oplus} < 3$). This sample contains 507 systems.
    \item \emph{ALL ISP}: an ``all inner small planet'' sample constructed from the concatenation of the SE and SN samples, plus additional systems in which all known transiting planets satisfy either the SE or SN radius criteria and contains at least one of each (i.e. \emph{SE+SN}). This sample contains 678 systems.
\end{enumerate}

By analyzing systems with known SEs and SNs separately, we can in principle measure the relative occurrence rates of OCs in each ISP subpopulation and assess whether/how OCs impact the compositions of ISPs. We do not analyze the SE+SN sample individually because it contains too few systems to robustly measure population trends (i.e. 37). We provide more details on this decision in Section~\ref{sec:Octopodes}.

\begin{figure}
    \includegraphics[width=\columnwidth]{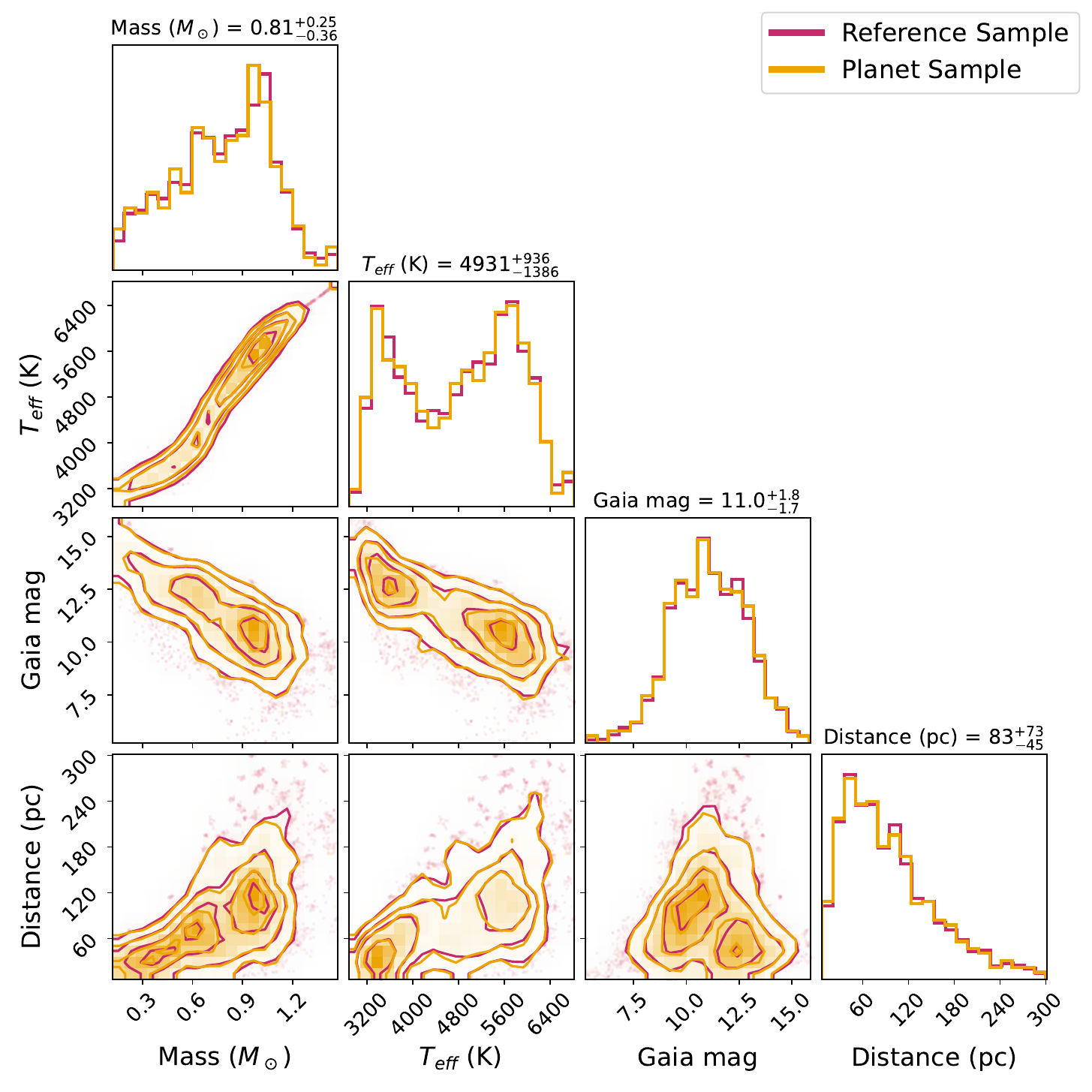}
    \caption{The joint stellar mass, effective temperature, Gaia magnitude, and distance distribution of host stars in our full planet sample (gold) and the corresponding reference sample (pink). The 1D histograms are normalized as we assign 10 reference stars for every planet host star. The median values of the reference sample are shown in the column titles with associated uncertainties from the 16$^{\rm th}$ and 84$^{\rm th}$ percentiles.}
    \label{fig:stars}
\end{figure}

For reference, Figure~\ref{fig:stars} depicts the distribution of host star parameters across our three planet samples.

\subsection{Reference Sample Selection}\label{sec:ref-sample}

\begin{figure*}
    \centering
    \includegraphics[width=1\linewidth]{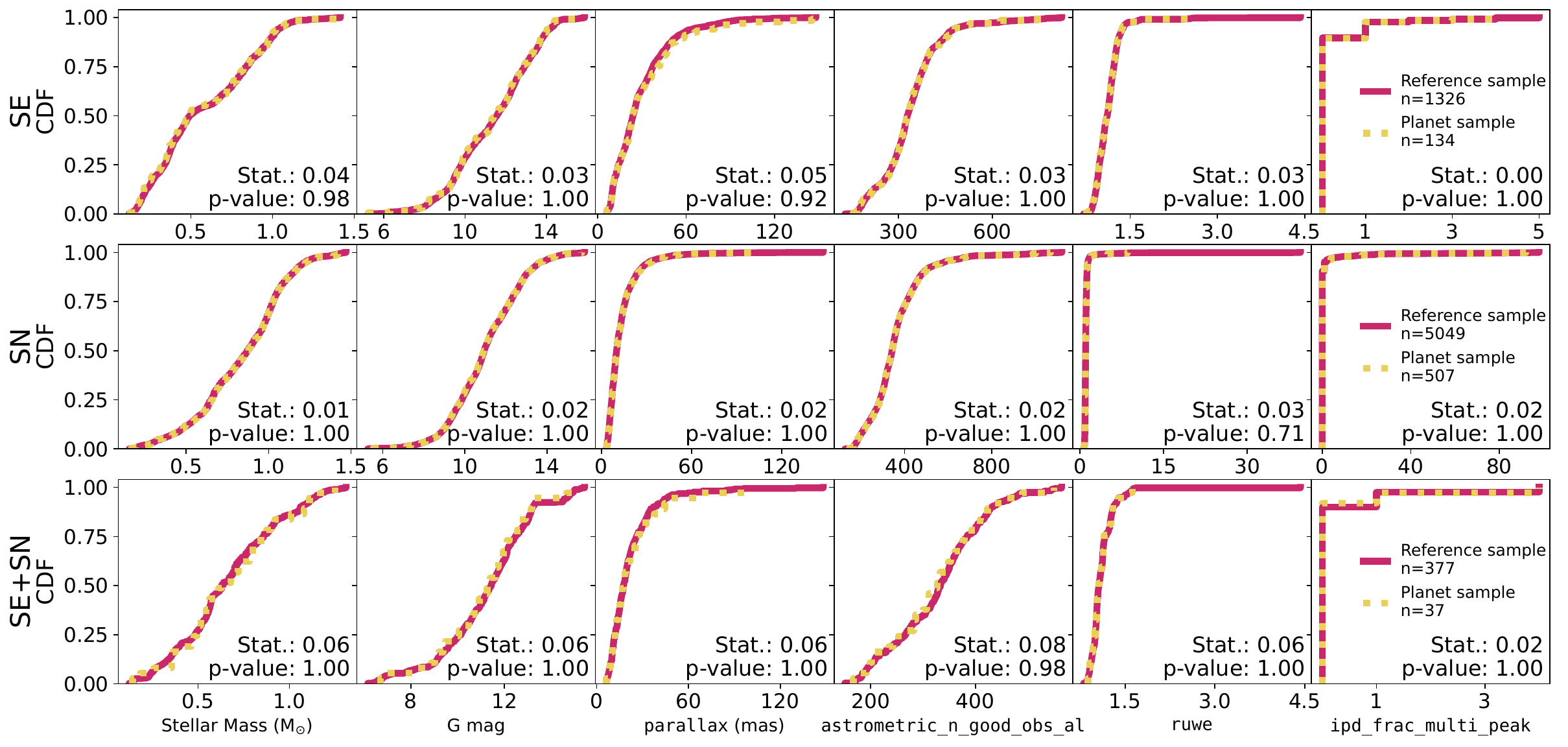}
    \caption{Cumulative distribution of stellar and astrometric parameters for the planet samples and their respective reference samples. The number of the reference samples reflect the number of stars that were included in the analysis.
    The KS statistic and p-value are shown at the bottom right corner of each subplot.}
    \label{fig:cdfs}
\end{figure*}

Calculating the relative occurrence of OCs in systems with ISPs, P(OC$|$ISP), versus in the field, P(OC), requires that we construct a reference stellar sample to each of our planet samples described in Section~\ref{sect:planet_sample}. We construct our reference samples by first identifying stellar parameters and observables that are known to affect the planet detection sensitivity in astrometry. We then search the TIC for stars 
with parameters that closely match the properties of the host stars in our planet samples. In this way, we posit that the astrometric detection sensitivity, while not calculated herein, is equivalent between each planet sample and its corresponding reference sample such that we can calculate \emph{relative} OC occurrence, i.e. P(OC$|$ISP)/P(OC), without knowledge of each sample's sensitivity function, which would be costly to calculate. 

We consider the following six parameters when constructing our reference stellar samples: stellar mass $M_\star$, raw Gaia magnitude $G$ (i.e. reddened), parallax $\varpi$, the number of good CCD observations in the along-scan (AL) direction of the source used in its astrometric solution (\texttt{astrometric\_n\_good\_obs\_al}), the renormalized unit weight error (\texttt{ruwe}), a measure of deviation of the astrometric solution assuming a single source, and the percentage of successful image parameter determination (IPD) windows with more than one PSF peak (\texttt{ipd\_frac\_multi\_peak}).
The first four parameters are known to affect astrometric detection sensitivity because they either control the planet-induced astrometric signal amplitude, the observational noise, or the sampling of the astrometric variability. The latter two parameters are stellar multiplicity diagnostics, where excess noise in the single-source astrometric solution with \texttt{ruwe} $\gtrsim 1.4$ is indicative of an unresolved binary system \citep{Lindegren2018}. Similarly, \texttt{ipd\_frac\_multi\_peak} is an indicator of partially resolved binaries with non-zero values indicative of multiple sources. Because the outcome of planet formation is known to vary with stellar multiplicity \citep[e.g.][]{Cieza2009, Kraus2012, Kraus2016, Moe2021}, we construct our planet and reference samples to retain information about the presence of binary stars such that we can marginalize over their astrophysical effects rather than attempt to clean our datasets of probable binaries based on incomplete, noisy data. 

The astrometric parameters necessary to match our planet and reference samples are taken from Gaia DR3, using the \texttt{dr2\_neighbourhood} auxiliary table to map between DR2 and DR3 IDs. 
We match the Gaia and TIC IDs at high accuracy 
by cross-matching J2015 RA and Dec against the G23H catalog \citep{Thompson2026}. This avoids issues when one TIC ID might point to more than one star in Gaia, as can occur when nearby sources are unresolved with TESS. 
%
We use the Nearest Neighbors search algorithm 
implemented in the Python library \texttt{scikit-learn.neighbors.NearestNeighbors} \citep{scikit-learn} to match each star in each of our planet samples with its ten nearest stars in the TIC that do not host any known planets in the six dimensional parameter space $\{ M_\star, G, \varpi,$ \texttt{astrometric\_n\_good\_obs\_al}, \texttt{ruwe}, \texttt{ipd\_frac\_multi\_peak}$\}$. We assign each of our planet samples its own reference sample, for a total of six stellar samples. 
One of these planet samples is the SE+SN sample, such that the concatenated ALL ISP sample is not itself matched using this methodology, but the three planet samples it contains are (i.e. SE, SN, and SE+SN). 

We show the cumulative distribution functions (CDF) for our planet samples and their respective reference samples across the six matched parameters in Figure~\ref{fig:cdfs}. It is clear that our approach using Nearest Neighbors yields highly consistent distributions for all six matched parameters and all planet/reference sample combinations. We compute two-sample Kolmogorov–Smirnov (KS) tests for each planet/reference sample pair to confirm that these samples are consistent with being drawn from the same parameter distributions, as intended. The results of our KS tests are reported in Table~\ref{tab:cdfs_limits}. Table~\ref{tab:cdfs_limits} also outlines the range of each parameter in each sample, conveying the level of agreement between our planet/reference sample pairs. However we note that due to our large sample sizes, small deviations in the CDFs can result in a lower-than-expected KS statistic ($\lesssim 1$). 
In Appendix~\ref{sec:cdf_appendix} we provide the CDFs of the resulting ALL ISP sample and resulting CDFs following cuts to our samples made during our forthcoming analysis (e.g., stellar mass and metallicity cuts). We note that due to mismatches between Gaia and TIC IDs, our reference samples can have slightly less than exactly $10\times$ the number of stars in its corresponding planet sample. However, these discrepancies are minor and are expected to have a negligible impact on our results. Our full sample contains 7,321 stars split between 678 and 6,643 stars across our planet and reference samples, respectively.

\begin{table}[]
    \centering
    \caption{Summary Table of Nearest Neighbor Matched Parameters.}
    \begin{tabular}{lccc}
        \hline
        \hline
        Parameter &  SE & SN & ALL ISP \\
\hline

        \multirow{3}{*}{Stellar Mass (M$_{\odot}$)} & 0.13-1.45 & 0.17-1.47 & 0.13-1.47 \\
 & 0.12-1.42 & 0.15-1.47 & 0.12-1.47 \\
& 0.98 & 1.00 & 1.00 \\
 
         \hline
        \multirow{3}{*}{G mag}  & 5.24-15.82 & 5.51-15.82 & 5.24-15.82 \\
        & 5.21-15.89 & 5.26-15.84 & 5.21-15.89 \\
        & 1.00 & 1.00 & 1.00 \\
         
         \hline
        \multirow{3}{*}{\texttt{parallax} (mas)}  & 5.92-153.08 & 3.34-153.08 & 3.34-153.08 \\
         & 5.84-148.13 & 3.32-147.61 & 3.32-148.13 \\
        & 0.92 & 0.92 & 1.00 \\

                 \hline
        \multirow{3}{*}{\parbox{2.5cm}{\raggedright\texttt{astrometric\_}\\
        \texttt{good\_obs\_al}}} & 155-784 & 147-1069 & 147-1069 \\
        & 129-822 & 127-1122 & 127-1122 \\
        & 1.00 & 1.00 & 0.96 \\
        
         \hline
        \multirow{3}{*}{\texttt{ruwe}}  & 0.75-2.53 & 0.75-37.83 & 0.75-37.83 \\
         & 0.70-4.43 & 0.65-39.55 & 0.65-39.55 \\
            & 1.00 & 0.67 & 0.72 \\

         \hline
        \multirow{3}{*}{\parbox{2.5cm}{\raggedright\texttt{ipd\_}\\
        \texttt{multi\_peak}}} & 0-4 & 0-92 & 0-92 \\
          & 0-5 & 0-98 & 0-98 \\
        & 1.00 & 1.00 & 1.00 \\

         \hline
        \multirow{2}{*}{{$N$ systems}}  & 134 & 507 & 683 \\
         & 1326 & 5049 & 6643 \\
                  \hline
    \end{tabular}
    \begin{tablenotes}
    \item Note -- Parameter ranges correspond to the full range of parameter values returned by the Nearest Neighbors match. For each parameter, we report the parameter range for the planet sample (top row), the corresponding reference sample (middle row), and the two-sample KS test statistic.
    \end{tablenotes}
    \label{tab:cdfs_limits}
\end{table}

\subsection{The Search for Cold Gas Giants in Calibrated Gaia DR2 \& DR3 Astrometry}

To search for massive companions around our sample stars, we use the orbit-modelling package \texttt{Octofitter} 
\citep{Thompson2023} to model the calibrated, absolute astrometry in the Gaia DR2-DR3-Hipparcos astrometric catalog \citep[G23H;][]{Thompson2026}.
This framework calibrates Gaia DR2 and DR3 \citep{gaia_collab_2018, gaia_collab_2023} absolute astrometry with Hipparcos data (when available) and reports it in the DR3 reference frame \citep{Thompson2026}. 
G23H reports calibrated Gaia DR2 and DR3 proper motion anomalies in the DR3 reference frame, in combination with DR3 astrometric excess noise derived from independently calculated noise estimates following the method of \citet{Kiefer2025} and Gaia radial velocity uncertainties from the \textit{paired} catalog \citep{paired}. 
G23H also makes use of the Hipparcos-Gaia Catalog of Accelerations \citep[HGCA;][]{hgca_dr3} for the proper motion and scaled position differences calibrated against the DR3 reference frame. We refer the reader to \citet{Thompson2026_supp} for additional details on the modelling.
G23H data enables detectability of unresolved massive companions that perturb the assumed single-object astrometric solutions, which when combined with our large sample of stars creates an ideal experiment to constrain the occurrence rate of OCs in systems with and without known ISPs.

In practice, only 88 out of 678 ISP hosts in our planet samples have Hipparcos data, with most M dwarf hosts missing from Hipparcos given the mission's bias towards earlier type stars. To retain our large planet sample that will allow us to measure relative OC occurrence for different types of ISPs and around different types of host stars, we elect to omit all Hipparcos observations from our companion search, thus maintaining a homogeneous astrometric dataset. 
Because we do not use any Hipparcos data, even when available, the observational baseline is shortened from 25 years to just 34 months, resulting in decreased companion detection sensitivity compared to the capabilities demonstrated in past work with the full G23H data \citep{Blakely2026, Doo2026, Thompson2026, Thompson2026_supp, Thompson2026_letter}. 
We run \texttt{Octofitter} on 7,321 systems initializing wide, uninformative, log-uniform priors on mass ratio $q=m/M_\star$, $\log{\mathcal{U}(10^{-5},1)}$, and orbital period $P$, $\log{\mathcal{U}(1\text{ day},10^4 \text{ years})}$. 
Less than 1\% of the stars in our samples lacked \textit{G23} calibrations. 
Each star is independently fit with four system models containing 0-3 companions, respectively, and sampled with non-reversible parallel tempering implemented in \texttt{Pigeons} \citep{pigeons}. 

We ran  \texttt{Octofitter} in the $\log q$--$\log P$ parameter space, which is a natural choice for astrometric sensitivity and allows us to ensure that detectable astrometric signals are contained within our priors. We then reweigh our resulting posteriors to $\log$ companion mass ($m$)--$\log$ semi-major axis ($a$) space given host stellar masses. With log uniform priors, $m$ and $a$ have the same density as $q$ and $P$, so we truncate our posteriors to remove draws that fall outside of the largest rectangular box we can fit within the parallelogram-shaped distribution of $m,a$ converted from $q,P$.
This effectively restricts our reweighted priors to $a=0.044-912.012$ AU and $m=0.107-416.869\, M_{\rm Jup}$.
We maintain a uniform number of samples for all chains by repeating draws to fill in the chains. Given that the sampler we use in the next step cycles through each chain many times, this does not affect our posteriors.

\subsection{Reconstructing the Distribution of Astrometric Companions with Bayesian Hierarchical Modelling}\label{sec:Octopodes}
Because i) we expect relative poor detection sensitivity of OCs from astrometry limited to Gaia DR2 and DR3\footnote{In comparison to Gaia DR2-DR3-Hipparcos \citep[e.g.][]{Thompson2026} and to expectations from Gaia DR4 epoch astrometry \citep{Lammers2026}.} and ii) because precise parameters on detected companions remain poorly constrained (e.g. with $\log{m}$ and $\log{a}$ ranges spanning multiple decades; \citealt{Thompson2026, Destriez2026}), we do not assign a detection/non-detection flag to each system and instead jointly fit the companion population using a Bayesian hierarchical model.
Specifically, we use the \texttt{Octopodes} framework developed to process per-system posteriors from \texttt{Octofitter}.
Here we provide a brief summary of \texttt{Octopodes}\footnote{https://github.com/Julia-Tempering/Octopodes.jl.git} but refer the reader to Appendix B of \citet{Thompson2026_letter} for a more detailed discussion.

\begin{figure}
    \centering
    \includegraphics[width=1.\linewidth]{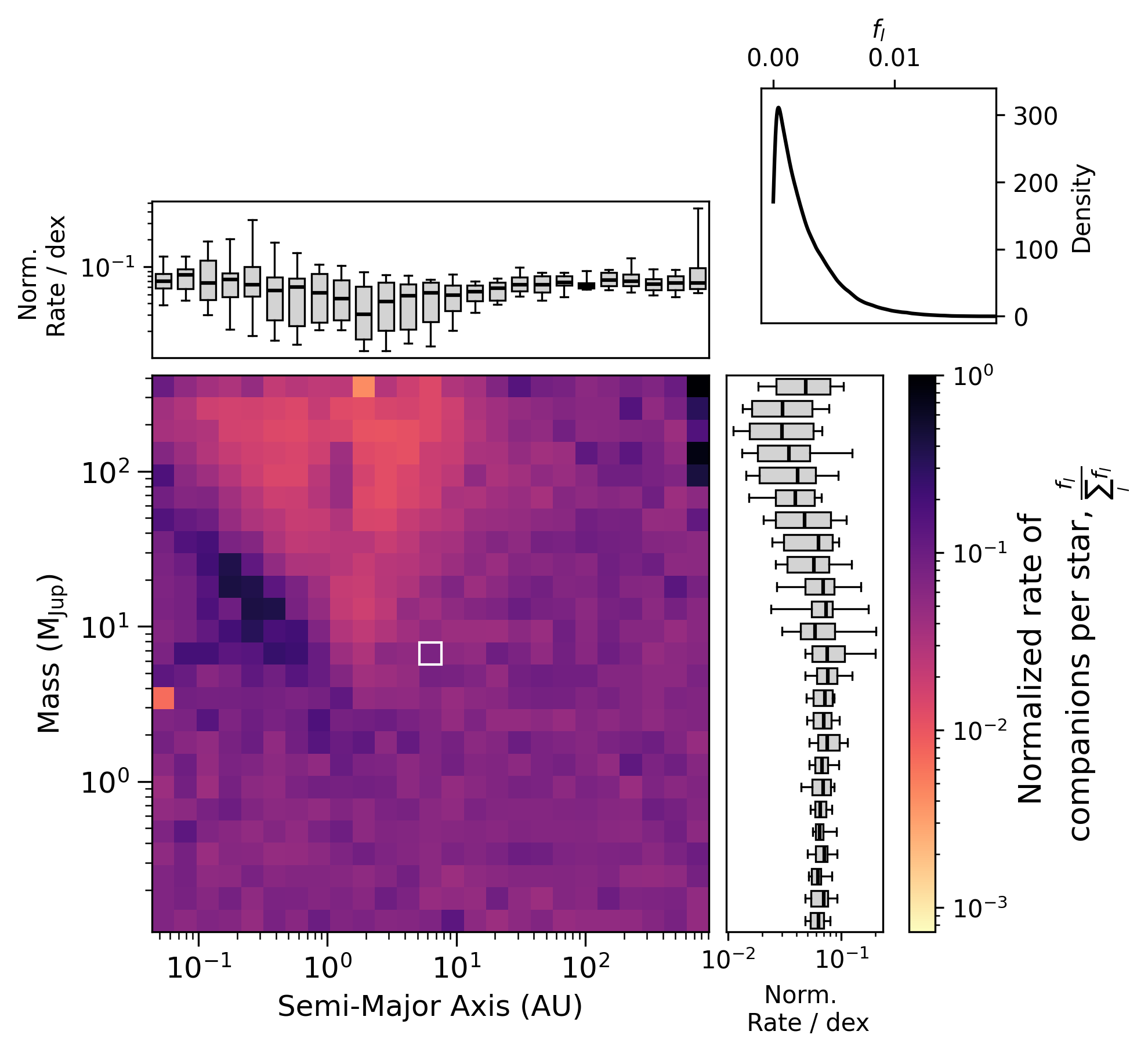}
    \caption{Example output from \texttt{Octopodes}, for the SN reference sample. The heat map shows the normalized rate of companions per star $f_l$, corresponding to the mean value of the distribution over 160,000 iterations (with 40,000 iterations discarded as burn in). The marginalized distributions per dex in semi-major axis and mass space are shown with box plots in the top and right sub-panels, respectively. The whiskers of the box plots denote the 5th and 95th percentiles of the marginalized distribution. The top right sub-panel shows an example zoom-in of the full posterior distribution of $f_l$ for one, randomly-chosen cell outlined in white on the heat map.  } 
    \label{fig:octopodes}
\end{figure}

The OC population model parameters of interest are denoted $\psi_n$, the distribution of the number of companions per star from $n=0,1,2,3$ and $\pi_l$, the distribution of (normalized) companion membership probability over our $\log{a}-\log{m}$ grid, where $l=1,\dots,L$ is the cell index in our $25\times 25$ $\log{a}-\log{m}$ grid (i.e. $L=625$). Both parameters are normalized to unity via $\sum_{n=0}^3\psi_n=1$ and $\sum_{l=1}^{625}\pi_l=1$.
With this hierarchical model, we measure the rate of companions per star per $\log{a},\log{m}$ cell: 
$f_l(\psi_n,\pi_l) = \pi_l \sum_{n=0}^3 n \psi_n$.
We note that when we calculate the total rate of companions over a select region of parameter space (see Section~\ref{sec:occurrence}), we substitute the notation $f_l$ with P(OC$|$sample), whereby `sample' represents the planet/star sample of interest (e.g., ALL ISP, SE, SN, etc.). Figure~\ref{fig:octopodes} shows an example output of the \texttt{Octopodes} population model for our SN reference sample.

Our Bayesian hierarchical modelling approach has two important advantages:
1) inferences about each system's rate of companions are informed by the overall sample, and 2) signals too weak to be detected individually are included in the population inferences, which would not be possible if we classified each posterior as a detection or non-detection.
Contrary to most Bayesian hierarchical models used for exoplanet demographics studies that use a two-step approach by first assigning a detection decision then using this as input for the Bayesian inference \citep[see discussion in][]{faria_inferring_2025}, \texttt{Octopodes}  models each system's data jointly and propagates its uncertainty over the rate of companions while simultaneously modelling detection completeness implicitly. 

In the population model, nearby stars are naturally weighted more heavily than distant stars because because they boast greater sensitivity even though we expect nearby stars to host the same OC population as more distant stars of the same spectral type and metallicity. Because our samples are calibrated in parallax, this weighting is consistent across planet/reference sample pairs such that it does not introduce any differential completeness between the two samples.
However, because low sample sizes will increase the uncertainty in where the population lies in $\log{a}-\log{m}$ space, not just the inferred occurrence rate in any cell, we only use the hierarchical model on samples containing more than 50 systems. We empirically determined this threshold by randomly reducing our original samples by half and re-running the population model to determine the impact of reduced sample size on the robustness of our population-level inferences. We found no discernible difference except for the SE planet sample whereby the sample size was reduced from 134 to 67 and the corresponding posterior shifted by 1.1$\sigma$; by far the largest shift observed.

We jointly fit our per-system posteriors in each of our planet and reference samples, treating each sample as its own population. 
We also consider additional populations by slicing our samples across stellar mass and metallicity, which allows us to probe distinct planet formation processes (see Section~\ref{sec:results}).
We initialize the sampler for 200,000 iterations and discard 20\% as burn-in. 

\subsection{Measuring Outer Companion Occurrence Rates}\label{sec:occurrence}
After computing our the companion rate $f_l$ for a given sample using our hierarchical population model (e.g., Figure~\ref{fig:octopodes}), we then integrate $f_l$ over a region of interest in the $\log{a}-\log{m}$ grid.
Much of the parameter space is dominated by the prior (i.e. the data have low sensitivity) so we restrict our occurrence rate inferences to the higher mass regime with higher sensitivity. 
We measure the occurrence rates of OCs in the range of $1-80\, M_{\rm Jup}$ and $1-20$ AU to probe cold, substellar companions as well as select GGs up to the limits set by our data.
Because these physical ranges do not fall exactly on the cell edges of our 25 $\times$ 25 $\log{a}-\log{m}$ grid, we assume a uniform distribution within a given cell and reweigh each cell proportionally to the fraction of its area contained within the integration region.
Where the posteriors are uninformative and revert to the prior, we apply a sensitivity cut. The resulting integration region is the union of our box between $1-80\, M_{\rm Jup}$ and $1-20$ AU, and our sensitivity mask.

We define sensitivity in each cell by the dependence of the posterior on the prior by calculating the derivative of the posterior mean 
with respect to the prior; $S\equiv \partial \langle f_l \rangle / \partial \mu$. In other words, we quantify the effect that changing the prior has on the recovered posterior.
To interpret $S$ after normalizing to the range $0-1$, a value of $S=0$ indicates no dependence on the prior whereas $S=1$ indicates full dependence on the prior (i.e. the data provide no information). 

To calculate OC occurrence within each planet sample, we mask cells whose sensitivity is greater than a chosen $S$ threshold defined using the planet sample. We then apply the same mask to the corresponding reference sample to facilitate a one-to-one comparison when calculating relative OC occurrence using Eq.~\ref{eq:rel}. 

We tested the full range of normalized sensitivity thresholds $S\in [0,1]$ in which we recalculated the relative OC occurrence distributions across increasing $S$ thresholds to assess the sensitivity of relative OC occurrence to our choice of threshold. We found that in our planet/star samples ALL ISP (683 systems), ALL ISP around M dwarfs (ALL ISP,MD; 184 systems), and ALL ISP around metal-poor stars (ALL ISP,MP; [M/H]$<0$; 288 systems), we identified sharp transitions in the relative OC occurrence at specific $S$ thresholds. Whereas the relative OC occurrence in our remaining samples ALL ISP around FGK dwarfs (ALL ISP,FGK; 494 systems) and ALL ISP around metal-rich stars (ALL ISP,MR; [M/H]$>0$; 341 systems) proved to be insensitive to our choice of $S$ threshold. 

\begin{figure}
    \centering
    \includegraphics[width=1\linewidth]{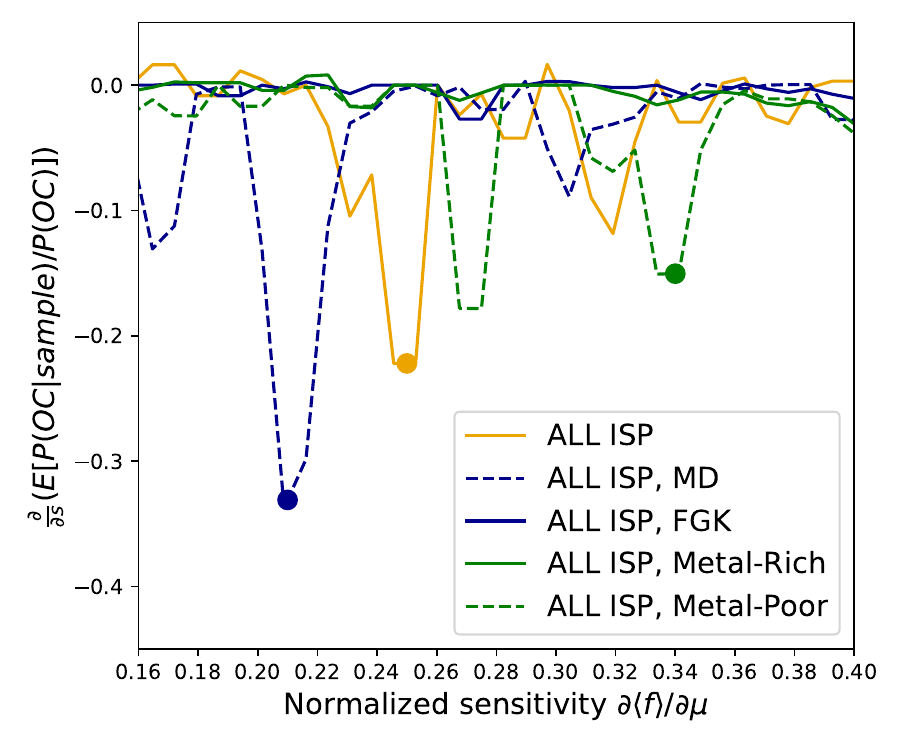}
    \caption{Fractional sensitivity thresholds tested for the sensitivity mask applied to the integration region when calculating the occurrence rates. Chosen thresholds are shown with the dots. Comparative samples (same color) share the same sensitivity threshold, which is determined based on the sample showing the most reactivity to the threshold.}
    \label{fig:gradient}
\end{figure}

These features are visualized in Figure~\ref{fig:gradient}, which for each of the five aforementioned samples, depicts the gradient with respect to $S$ of the expectation value of the relative OC occurrence (i.e. E[P(OC$|$sample)/P(OC)]). When the gradient is shallow and near zero, the relative OC occurrence is largely insensitive to our choice of $S$ threshold, however, Figure~\ref{fig:gradient} reveals that this is not always the case for certain planet/star samples. We note that Figure~\ref{fig:gradient} depicts a zoomed in region of the full $S$ range (i.e. $S\in [0.16,0.40]$ where the relevant gradient structures are manifested. 
The samples ``ALL ISP'', ``ALL ISP,MD'', and ``ALL ISP,MP'' have at least one sharp drop (i.e. a large negative gradient) in relative OC occurrence indicating that as the $S$ threshold is increased and consequently more of the $\log{a}-\log{m}$ parameter space is added to the integration region, the relative OC occurrence changes sharply. To select the most appropriate $S$ threshold, we adopt the value of $S$ that corresponds to the last dip in the gradient curve to ensure that our forthcoming inferences on OC occurrence are not based on very small samples that occur when the $S$ threshold is low (i.e. the integration region is highly restricted). Following these gradient dips, the gradient reverts back to small values near zero indicating that the relative OC occurrence is now independent of our sensitivity mask. Our chosen $S$ thresholds for ``ALL ISP'', ``ALL ISP,MD'', and ``ALL ISP,MP'' are 0.25, 0.21, and 0.34, respectively. To enable direct comparison between certain planet/star samples, we apply the same ALL ISP threshold of 0.25 to our SE and SN samples. Similarly, we adopt the same threshold of 0.21 for the ``ALL ISP,MD'' and ``ALL ISP,FGK'' samples, and the same threshold of 0.34 for the ``ALL ISP,MP'' and ``ALL ISP,MR'' samples.

\begin{figure*}
    \centering
    \includegraphics[width=1.\linewidth]{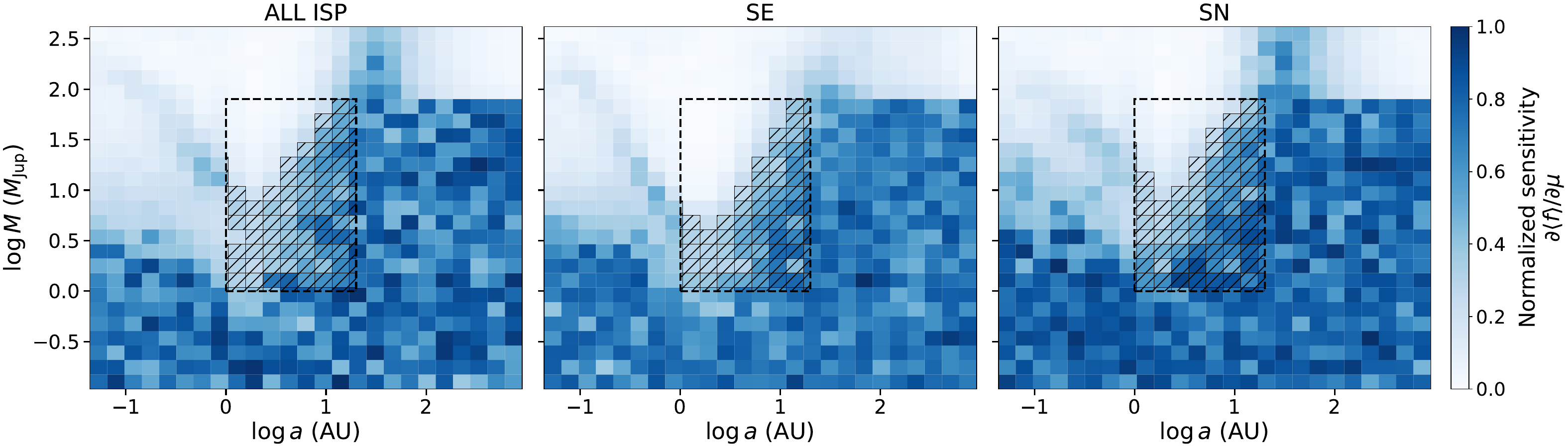}
    \caption{Sensitivity maps for our ALL ISP, SE, and SN planet samples. Sensitivity is calculated as the normalized derivative of the average posterior $\langle f_l \rangle$ with respect to the prior $\mu$. Regions where the normalized derivative is large correspond to where the posterior is prior-dominated and the detection sensitivity to astrometric companions is correspondingly low. The dashed rectangle shows our unmasked integration region between $1-20$ AU and $1-80\, M_{\rm Jup}$. Hatching corresponds to the cells excluded from the analysis due to exceeding our sensitivity threshold, in these cases, corresponding to the $25^{\rm th}$ percentile.  }
    \label{fig:sensitivity}
\end{figure*}

We show example sensitivity maps for our ALL ISP, SE and SN planet samples in Figure \ref{fig:sensitivity}.
 The rectangle corresponds to the  $1-20$ AU  and $1-80\, M_{\rm Jup}$ ranges, whereas the hatched-out cells correspond to those excluded by our sensitivity mask.
In Section \ref{sec:results}, we report occurrence rates measured using our sensitivity mask. For comparison and completeness, in Section~\ref{subsec:sensitivity-agnostic} we also report the occurrence rates integrated over our full $\log{a}$--$\log{m}$ range, thus including regions of the parameter space that are prior-dominated.

\section{Results} \label{sec:results}

\subsection{Relative Occurrence Rates of Outer Companions in Systems with Inner Small Planets}

\begin{figure}[h!]
    \centering
    \includegraphics[width=1.\linewidth]{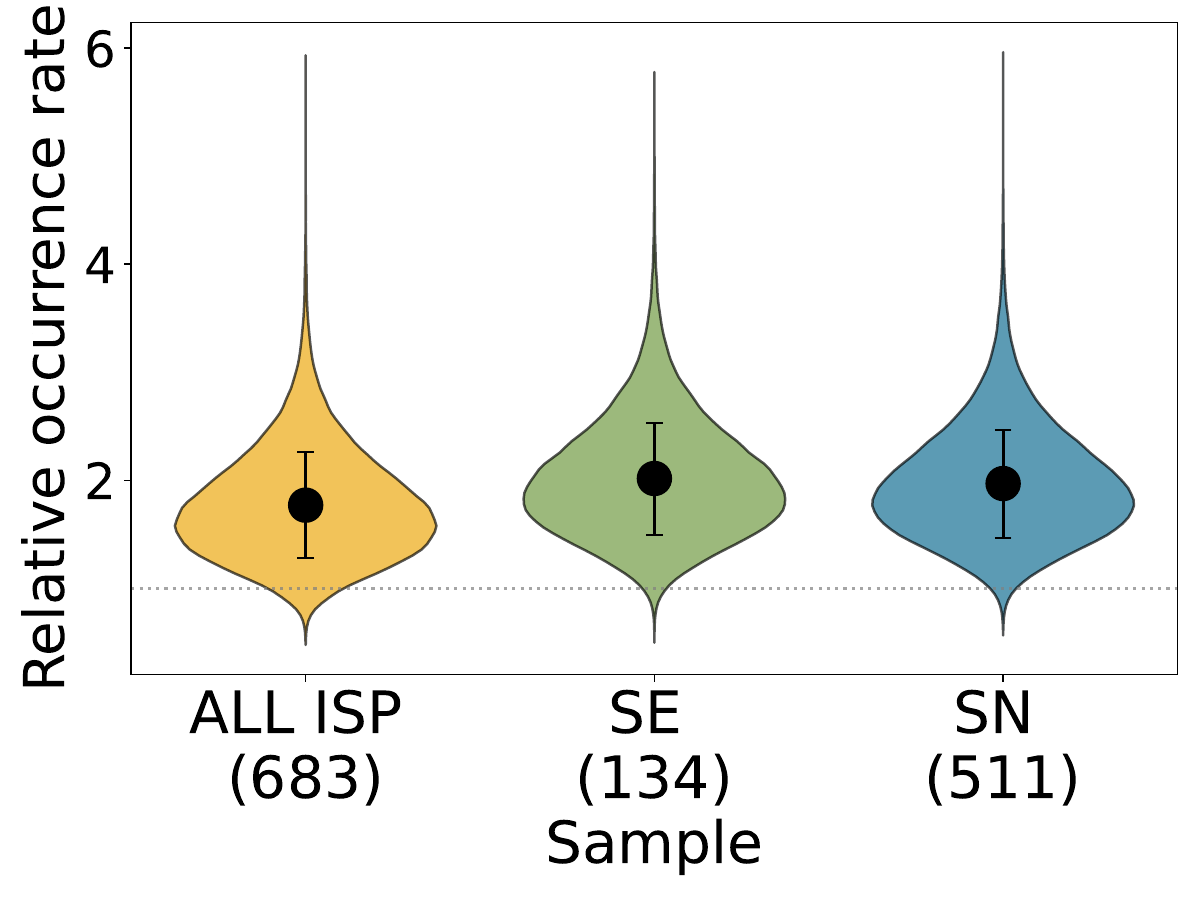}
    \caption{Relative occurrence rates of outer companions P(OC$|$sample)/P(OC) for three inner planet samples: ALL ISP, SE, and SN. The error bars depict the $16-84\%$ confidence interval with the central dot depicting the expectation value.}
    \label{fig:ratio_violin_main}
\end{figure}

Figure~\ref{fig:ratio_violin_main} shows our distributions of the relative occurrence rates of OCs for our three inner planet samples: ALL ISP, SE, and SN. These relative occurrence rates are the ratio of the conditional occurrence of OCs in systems with select ISPs to the occurrence of OCs in the calibrated reference sample of matched stars in the field that lack known planets; P(OC$|$ISP)/P(OC). Recall from Section~\ref{sec:methods} that we do not construct sensitivity corrections for our individual planet and reference samples such that we can only report robust relative occurrences, not absolute occurrences.

We calculate the expectation values of each relative occurrence rate distribution, which range from $1.77-2.02$ across the three planet samples in Figure~\ref{fig:ratio_violin_main}. These results indicate strong evidence for an excess of OCs in systems with ISPs by a factor of $\sim 2$, regardless of whether the inner planets are super-Earths or sub-Neptunes, or both. 
We also find that the relative OC occurrence rates are $>1$ with $>96.8\%$ confidence, further indicating strong evidence of OC excess in ISP systems and validating consensus results from RV surveys. 
Explicitly, we find that P(OC$|$ALL ISP)/P(OC) $>1$ with 96.8\% confidence (i.e. $1.85\sigma$), and the smaller planet samples P(OC$|$SE)/P(OC) and P(OC$|$SN)/P(OC) are $>1$ with 99.3\% confidence (i.e. $2.46\sigma$) and 98.2\% confidence (i.e. $2.10\sigma$), respectively.

Each ratio's expectation value and its $\pm1\sigma$ uncertainty from the $16^{\rm th}$ and $84^{\rm th}$ percentiles are summarized in Table~\ref{tab:main_results}. 

\renewcommand{\arraystretch}{1.35}
\begin{table}[]
    \centering
    \caption{Relative occurrence rates of outer companions in systems hosting inner small planets.} 
    \begin{tabular}{lccc}
        \hline
        \hline
 Host Stars&  $\frac{\textrm{P(OC$|$ALL ISP)}}{\textrm{P(OC)}}$ & $\frac{\textrm{P(OC$|$SE)}}{\textrm{P(OC)}}$   & $\frac{\textrm{P(OC$|$SN)}}{\textrm{P(OC)}}$\\
         \hline
All stars & $1.77\pm{0.49}$ & $2.02\pm{0.52}$ & $1.97\pm{0.50}$ \\
M dwarfs & $2.44\pm{0.64}$ & $1.81\pm{0.51}$ & $2.01\pm{0.55}$ \\
FGK & $1.73\pm{0.45}$ & $1.61\pm{0.52}$ & $1.59\pm{0.45}$ \\
\hline
    \end{tabular}
    \label{tab:main_results}
\end{table}

\subsection{Relative Outer Companion Occurrence Across Stellar Masses} \label{sec:res_mass}

\begin{figure*}
    \centering
    \includegraphics[width=1.\linewidth]{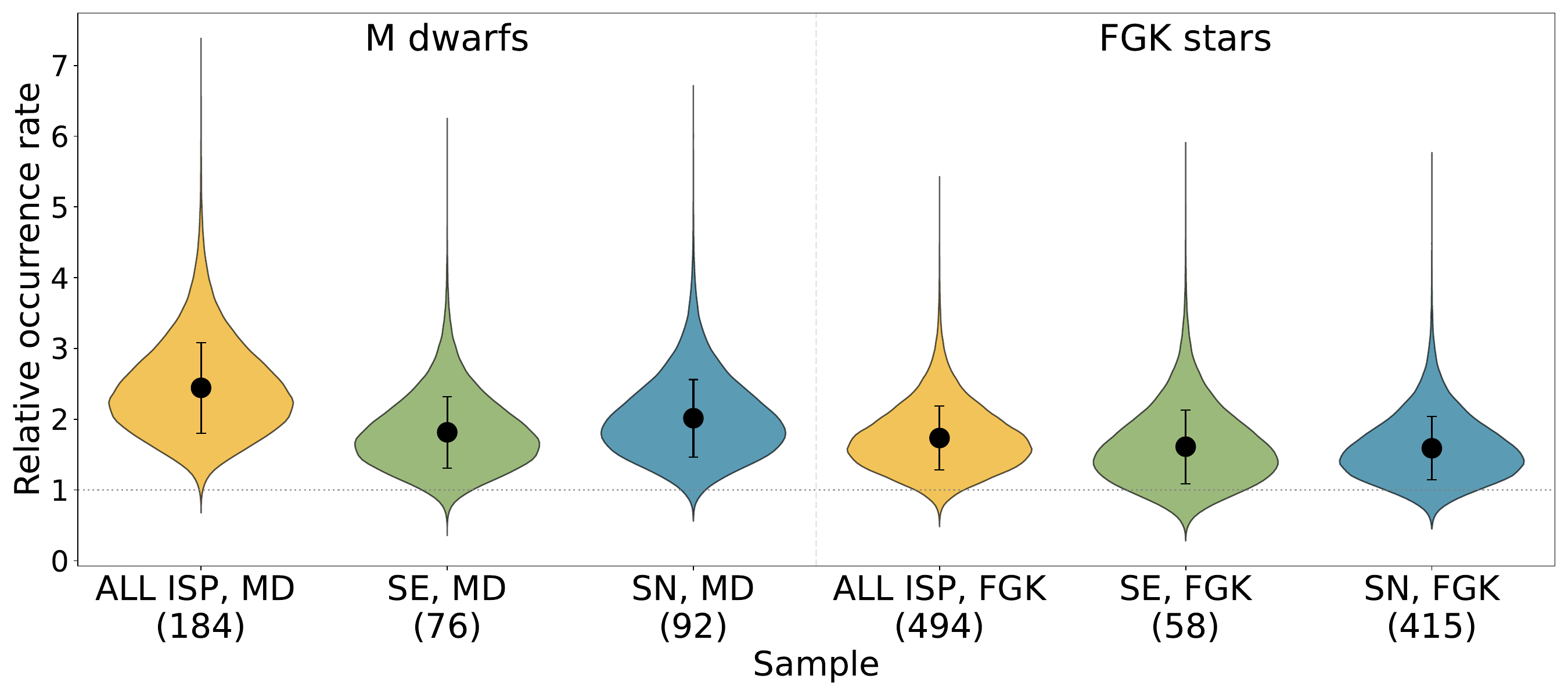}
    \caption{Relative occurrence rates of outer companions P(OC$|$sample)/P(OC) for six inner planet samples: ALL ISP, SE, and SN with either M dwarf hosts (left) or FGK dwarf hosts (right). The error bars depict the 16-84\% confidence interval with the central dot depicting the expectation value. 
    }
    \label{fig:ratio_violin_stellar_mass}
\end{figure*}

Here we investigate how the relative occurrence rates vary across coarse stellar mass bins containing M and FGK dwarfs, respectively. We classify stars based on stellar masses with M dwarfs defined as $M_\star\leq0.6\, M_\odot$ while FGK dwarfs have $M_\star>0.6\, M_\odot$. We note that our full planet sample is biased towards FGK dwarfs, representing 72\% of our full planet sample (c.f. Figure~\ref{fig:stars}). This is a direct consequence of the TESS sensitivity function from which we draw our planet samples and is independent of our astrometric data.

Figure~\ref{fig:ratio_violin_stellar_mass} reveals that, as in the case that marginalizes over host star masses in Figure~\ref{fig:ratio_violin_main}, the relative occurrence rates of OCs has a high probability of exceeding unity for both M and FGK dwarf systems with ISPs. We similarly observe an excess of OCs in systems hosting either SEs or SNs across spectral types.
These relative OC occurrence rates are also included in Table~\ref{tab:main_results}.

The occurrence rates of OCs in systems hosting ISPs are consistent for both M and FGK dwarfs, with 
P(OC$|$ALL ISP,MD)/P(OC) being slightly elevated by $1.41\pm0.78\times$ compared to P(OC$|$ALL ISP,FGK)/P(OC). 
This is despite our FGK planet sample for ALL ISPs being greater than the corresponding sample size around M dwarfs by $2.7\times$, which translates into more accurate constraints on OC relative occurrence \citep{Thompson2026_letter}.
We caution that our SE samples around M and FGK dwarfs and our SN sample around M dwarfs contain fewer than 100 systems, severely reducing the sensitivity of our hierarchical population model and thus limiting the reliability of any population inferences. 

Recall that our reference sample selection was not explicitly performed on planet samples with specifically M dwarf or FGK dwarf hosts, but results from slicing our calibrated samples. Therefore, the level of consistency between the matched parameters in these planet/reference sample pairs is not guaranteed to be high and may show some deviations. We reemphasize that large deviations between the matched stellar and astrometric parameters of a planet/reference sample pairing could compromise the equivalency of each sample's astrometric detection sensitivity to OCs, thus introducing a bias in our inferred relative OC occurrence rates. In 
Appendix~\ref{sec:cdf_appendix} we show the CDFs of the matched parameters for our planet and reference samples around M and FGK dwarfs and report the number of planetary systems in each of these sample slices. We find that there remains strong agreement between all planet/reference sample pairs while some minor deviations do exist that we acknowledge and discuss in Appendix~\ref{sec:cdf_appendix}.

\subsection{Relative Outer Companion Occurrence Across Stellar Metallicity} \label{sec:res_met}

\renewcommand{\arraystretch}{1.35}
\begin{table*}[]
    \centering
    \caption{Relative occurrence rates of outer companions in systems hosting inner small planets around metal-rich ([M/H]$>0$) and metal-poor ([M/H]$<0$) stars. Occurrence rates for the metal-rich SE and ISP, MD samples are not shown because their sample size is too small to be included in the analysis.} 
    \begin{tabular}{lccccc}

        \hline
 Metallicity &  $\frac{\textrm{P(OC$|$ALL ISP)}}{\textrm{P(OC)}}$ & $\frac{\textrm{P(OC$|$SE)}}{\textrm{P(OC)}}$   & $\frac{\textrm{P(OC$|$SN)}}{\textrm{P(OC)}}$ & $\frac{\textrm{P(OC$|$ALL ISP,MD)}}{\textrm{P(OC)}}$ & $\frac{\textrm{P(OC$|$ALL ISP,FGK)}}{\textrm{P(OC)}}$ \\
         \hline
Metal-Rich & $1.58\pm{0.39}$ & $-$ & $1.35\pm{0.35}$ & $-$ & $1.31\pm{0.33}$ \\
Metal-Poor & $2.10\pm{0.52}$ & $1.57\pm{0.41}$ & $1.55\pm{0.36}$ & $2.01\pm{0.48}$ & $1.34\pm{0.33}$ \\
\hline
    \end{tabular}
    \label{tab:metallicity}
\end{table*}

We obtained metallicities for 5,383 stars in our full stellar sample (N = 7,321) measured by the Gaia mission, specifically by the General Stellar Parametrizer-spectroscopy module  \citep[GSP-Spec;][]{recioblanco2023}. These metallicities are measured solely from spectra obtained with Gaia's Radial Velocity Spectrometer \citep[RVS;][]{Cropper2018}, and are deliberately uncalibrated. We thus calibrated the metallicities to mitigate discrepancies between the GSP-Spec metallicities and well-benchmarked literature values following \citep{recioblanco2023}. In order to perform the calibration, we used uncalibrated GSP-Spec surface gravities (log\textit{g}, c.f. Section 9.1.1 and Table 3 in \citet{recioblanco2023}). We only selected metallicities where the first 13 quality flags are $\leq 1$ in order to match the medium-quality sample reported in \citet{recioblanco2023}. Our resulting metallicities have a median value of +0.1 dex and a standard deviation of 0.2 dex. 

\begin{figure}
    \centering
    \includegraphics[width=\linewidth]{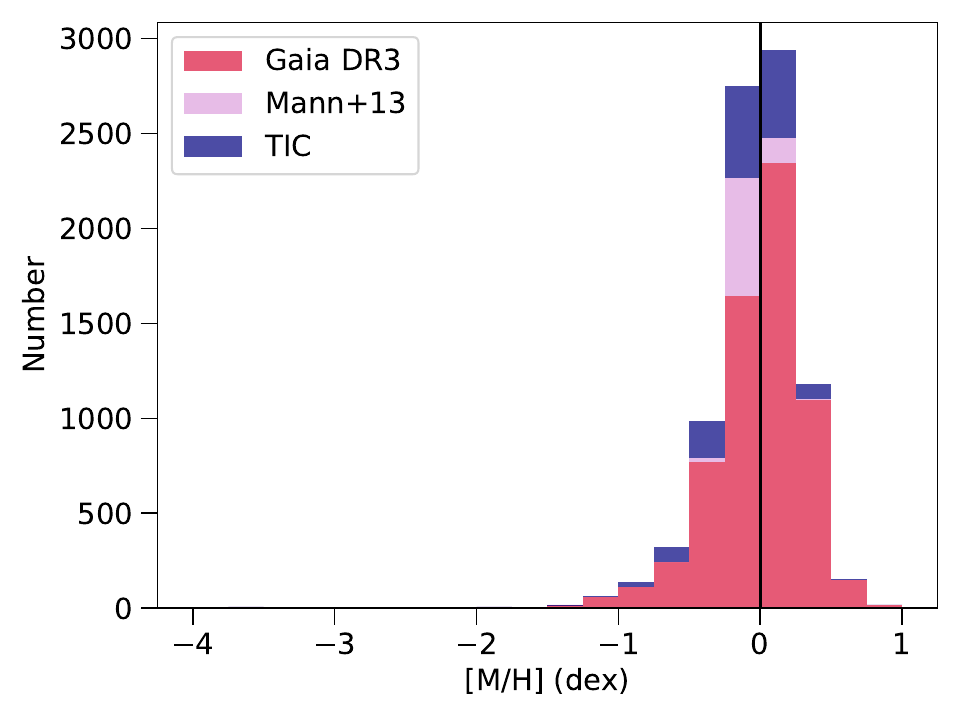}
    \caption{The distribution of [M/H] metallicities for 6,794 stars in our full stellar sample of 7,321. We retrieved [M/H] values for 5,383 stars calculated using Gaia DR3 \citep[pink,][]{recioblanco2023}, 779 stars using the \citet{Mann2013} empirical relations (light purple), and 632 stars with metallicities reported in the TIC \citep[dark blue,][]{Stassun2019}.}
    \label{fig:mhdistribution}
\end{figure}

Within our stellar sample, the GSP-Spec metallicities are biased towards more massive stars, with no stars cooler than 3500 K appearing in the catalogue. To increase our sample size, we calculate additional metallicities using the empirical color-metallicity relations presented in \citet{Mann2013}. These relations are calibrated using known FGK-M binaries, and we apply them to stars in our sample that do not have GSP-Spec metallicities and that have $(V-K) > 3.8$. We calculate [M/H] for 779 stars in our full stellar sample using this method. Finally, we include metallicity measurements for any of the remaining stars from the TIC \citep{Stassun2019}, where available (632 stars). In total, we obtain metallicity measurements for 6,794 stars (i.e. 93$\%$) in our sample. We show the distribution of metallicities from each source in Figure~\ref{fig:mhdistribution}.
With these heterogeneous [M/H] values, we proceed by classifying our stars as either metal-rich ([M/H] $>0$ dex) or metal-poor ([M/H] $\leq0$ dex). We then calculate the relative occurrence of OCs in these coarse metallicity bins.

\begin{figure*}
    \centering
    \includegraphics[width=1.\linewidth]{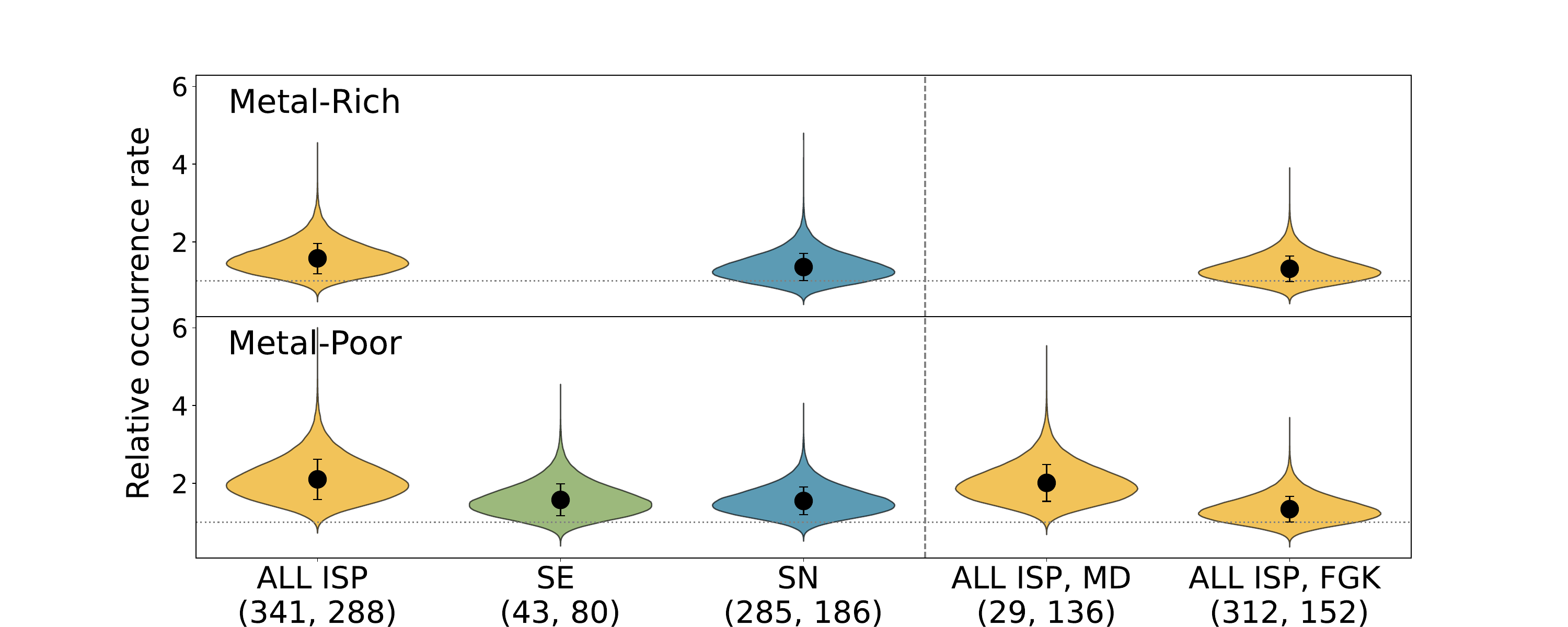}
    \caption{Relative occurrence rates of outer companions P(OC$|$sample)/P(OC) for five inner planet samples around each stellar sample of metal-rich (top; [M/H]$>0$) and metal-poor stars (bottom; [M/H]$<0$): ALL ISP, SE, SN across all host spectral types, plus ISPs around M and FGK dwarfs. The error bars depict in $16-84\%$ confidence interval with the central dot depicting the expectation value. The number of systems in each corresponding planet sample is written below the sample label for metal-rich, metal-poor respectively. Samples with fewer than 50 systems are not included in our analysis (i.e. SE, Metal-Rich and ALL ISP, MD, Metal-Poor.
    } 
    \label{fig:ratio_violin_metallicity}
\end{figure*}

Figure~\ref{fig:ratio_violin_metallicity} and Table~\ref{tab:metallicity} report our relative occurrence rates of OCs for the ALL ISP, SE, and SN planet samples in both our metal-rich and metal-poor samples. We also calculate P(OC$|$ISP)/P(OC) for our M dwarf and FGK dwarf samples for both metal-rich and metal-poor stars when at least 50 systems are present in the planet sample  SNs may occur less frequently around metal-poor \citep{Petigura2018}, although other work has suggested that both SEs and SNs around FGK dwarfs are agnostic to stellar metallicity \citep{Kutra2021}. For M dwarfs, SNs appear to preferentially form around metal-rich host stars \citep{Turtelboom2026}. These results highlight the interplay between stellar mass and metallicity in planet formation. Because we limit our analysis to samples containing more than 50 systems due to considerations of limited OC population inferences, we do not report the occurrence rates for the SE and SN M dwarf and FGK dwarf samples. In Appendix~\ref{sec:cdf_appendix} we show the CDFs of matched parameters for the ten aforementioned samples and reports the number of systems in each.

Inspection of Figure~\ref{fig:ratio_violin_metallicity} reveals that in all metal-rich and metal-poor samples, the relative occurrence rates of OCs are comparable to our previous data slices that marginalized over stellar metallicity, with some occurrence rates being closer to unity (c.f. Figures~\ref{fig:ratio_violin_main} and \ref{fig:ratio_violin_stellar_mass}). 
The metal-poor samples show ratios $>1$ with 99.7\% ($2.8\sigma$) confidence for both the ALL ISP and ALL ISP, MD samples, while the SE and SN ratios are $>1$ with 93.6\% ($1.4\sigma$) and 95.7\% ($1.7\sigma$) confidence, respectively. 
The metal-rich samples show ratios $>1$ with 95.1\% confidence ($1.7\sigma$) for the ALL ISP, while the SE ratio is consistent with unity at $1\sigma$. 
ALL ISP, FGK samples are consistent with unity at $1\sigma$ for both the metal-rich and metal-poor samples.
This suggests that metallicity is not a main driver of the conditional occurrence of OCs in systems with ISPs for the types of OCs that we are sensitive to in this study, although the interpretation might be more subtle. We reserve a discussion of this result for Section~\ref{sec:dis_metallicity}

\subsection{Sensitivity-Agnostic Measurements}\label{subsec:sensitivity-agnostic}

\renewcommand{\arraystretch}{1.35}
\begin{table}[]
    \centering
    \begin{tabular}{lccc}

        \hline
        \hline

Host Stars &  $\frac{\textrm{P(OC$|$ISP)}}{\textrm{P(OC)}}$ & $\frac{\textrm{P(OC$|$SE)}}{\textrm{P(OC)}}$   & $\frac{\textrm{P(OC$|$SN)}}{\textrm{P(OC)}}$\\
         \hline
All stars & $1.24\pm0.19$ & $1.17\pm0.24$ & $1.07\pm0.17$ \\
M dwarfs & $1.50\pm0.24$ & $1.10\pm0.24$ & $1.17\pm0.21$ \\
FGK & $1.11\pm0.19$ & $1.38\pm0.40$ & $1.05\pm0.18$ \\
\hline
    \end{tabular}
    \caption{Sensitivity agnostics conditional occurrence rate for outer companions in systems hosting inner small planets, inner super-Earths, inner sub-Neptunes and both super-Earths and sub-Neptunes, relative to the field occurrence of outer companions. 
    }
    
    \label{tab:sensitivity-agnostic}
\end{table}
Because of the sensitivity threshold we impose, our effective integration region does not correspond to the  1-20 AU  and 1-80 $M_{\rm Jup}$ physical range we are interested in, as shown in Figure \ref{fig:sensitivity}. We therefore report the relative conditional occurrence rates of OCs across the full physicial range (i.e. independent of the sensitivity threshold) in Table \ref{tab:sensitivity-agnostic} with the caveat that these values are likely noise- and/or prior-dominated. 
The ratio we present are all mostly consistent with 1 or slightly $>1$ at $\sim1\sigma$, as expected for prior-dominated results.

\subsection{Binaries}
The outcome of planet formation differs between single stars versus multi-star systems \citep{Moe2021,Venturini2026}.
Instead of attempting to clean our planet samples of binaries, 
we constructed our planet and reference samples by matching the distributions of the Gaia stellar multiplicity indicators \texttt{ruwe} and \texttt{ipd\_frac\_multi\_peak}, thus marginalizing over stellar multiplicity effects.  
However, as a sanity check, we repeat our calculations of relative OC occurrence rates on a strict slice of our sample that excludes probable binaries. We do this by rejecting all stars with \texttt{ruwe} $>1.4$ and \texttt{ipd\_frac\_multi\_peak} $\neq 0$. 
We only do this experiment for our full star sample without splitting by stellar mass or metallicity.
The resulting planet samples retain a total of 4,708 stars. 
We recover the consistent results to 0.40$\sigma$.

\section{Discussion} \label{sec:Discussion}
\subsection{Comparison to Previous Radial Velocity Studies}
This work provides the first astrometric confirmation of the excess of OCs in systems hosting ISPs previously reported by RV surveys \citep{Bryan2016,ZhuWu2018,Bryan2019,Bonomo2023, Bryan2024, Zhu2024, Bonomo2025, Bryan2025, VanZandt2025, DiChang2026, Quanyi2026, Ruggieri2026, Tang2026}.
Our study is highly complementary to RV studies given the uniqueness of our dataset (i.e. Gaia DR2+DR3), our homogeneous search and characterizeation of OC occurrence using a hierarchical Bayesian model, and because we have analyzed an increased sample size by an order of magnitude on average. The largest RV study by far included 1,167 solar-type stars \citep{Bonomo2025}.

Contrary to RV studies, we do not measure an absolute conditional occurrence rate of GGs P(GG$|$ISP) and then compare it to literature values of GG occurrence rates in the field \citep[P(GG);][]{Bryan2019,Bryan2024, Bryan2025, Bonomo2023, Bonomo2025}.
We instead report the relative conditional occurrence rate of OCs with respect to the field occurrence, P(OC$|$ISP)/P(OC), where we measure both occurrence rates modulo completeness corrections. Our planet and reference stellar samples are constructed to ensure that they share a common completeness such that we can report robust \emph{relative} occurrence rates. In other words, we do not rely on a comparison with a different, inhomogenous studies of field OC occurrence, but we also unable to measure absolute conditional occurrence rates that can be interpreted on their own. 

We also note that previous RV studies are sensitive to GGs, typically down to 1, or sometimes 0.1 $M_{\rm Jup}$, whereas we are most sensitive (i.e. not prior dominated) to more massive OCs $\gtrsim 10\, M_{\rm Jup}$, such that we are biased to probing substellar OCs. We therefore measure relative OC occurrence from 1-80 $M_{\rm Jup}$ range and $1-20$ AU, although we note that we are not sufficiently sensitive over this entire range such that our effective region in mass-semi-amjor axis space is restricted by insensitivity to prior assumptions (c.f. Figure~\ref{fig:sensitivity}). The outer limit on the semi-major axis sensitivity is limited to $\sim 20$ AU because the longest baseline in our data comes from the Gaia DR3 excess noise, which spans the DR3 mission duration of 34 months. 
We also do not search for OCs within 1 AU, as these objects would be very close to the inner systems, leading to destabilizing dynamical interactions. 

Our large planet sample sizes of 134 SE, plus 507 SN systems, allows us to robustly calculate relative OC occurrence rates in systems hosting SEs or SNs separately.
As shown in Figure \ref{fig:ratio_violin_main}, we find an excess of OCs in systems hosting ISPs, regardless of small planet type. Through an analysis of 43 systems, heavily biased to SNs, 
\citet{Tang2026} sought to measure P(GG$|$SE) and P(GG$|$SN) using different metrics to classify SEs and SNs, notably bulk density, envelope mass fraction, and core mass, all estimated from archival mass and radius measurements. \citet{Tang2026} find no significant differences between their subsamples, but hint that, for metal-rich systems, GGs may be more strongly preferred within SN systems compared to SE systems. This finding is supported by \cite{DiChang2026} who reported a similar conclusion based on a sample of 28 SEs and 52 SNs. With our planet sample size that is at least $6.2\times$ larger, and an order of magnitude larger than these RV studies on average, we robustly confirm that OC occurrence is elevated in systems hosting SEs and SNs compared to in the field, but find no significant difference between the excess of OCs in inner SE or SN systems (c.f. Figure~\ref{fig:ratio_violin_main}).

\citet{Bonomo2025} caveat their results, (and in effect those of \citet{Bryan2024, Bryan2025}), by saying that their computed completeness values are optimistic because they do not take into account uncorrelated jitters and ignore a small amount of RV trends that are not attributed to a confirmed GGs for the calculation of the occurrence rates, resulting in possibly underestimated occurrence rates.
Also, due to RV surveys being limited to the characterization of minimum masses $M_{\rm OC}\sin{i}$, a subset of the RV signals studied are likely induced by substellar companions that fall within the mass range that we probe in this work. 

\subsection{Implications for ISP Formation} \label{sec:formation}
Multiple planet formation models predict that OCs can inhibit the accretion of significant water mass fractions (WMFs) onto growing protoplanets by partially blocking the inward drift of icy pebbles from the outer disk, hence favoring the formation dry SEs \citep[e.g.,][]{Bitsch2019, Lambrechts2019, bitsch2021, Mulders2021}.
Herein, we observe a strong positive correlation between OC and ISP occurrence for both inner SEs and inner SNs (c.f. Figure~\ref{fig:ratio_violin_main}), with no significant difference in relative between the two subpopulations of ISPs. This suggests that OCs do not inhibit the formation of ISPs and that OCs do not play a dominant role in setting the bulk compositions of ISPs (i.e. their WMFs).

Spectroscopic observations of protoplanetary disks with JWST-MIRI have shown that the inner regions of smooth, compact (i.e. ringless) disks are rich in water compared to extended, ringed disks \citep{Banzatti2023}. 
This finding verifies that pressure bumps colocated with disk gaps--like those carved out by massive OCs--do inhibit the inward migration of icy pebbles even if they do not provide an impenetrable barrier \citep{Huang2025, Sandhaus2025, Houge2026}. Given this validation, we expect that if ISPs formed in situ then OCs and ISPs should exhibit an anti-correlation as OC-induced gaps would limit pebble delivery to in the inner system needed to form SEs and SNs of up to a few $M_\oplus$. This prediction is inconsistent with our measurement of a positive correlation between OCs and ISPs, which we reconcile with the results of \citet{Banzatti2023} by interpreting this as evidence against in-situ formation. Instead, we suggest that ISPs formed farther out in the disk than their present-day locations within $30$ days, although not necessarily beyond the snow line and certainly not beyond the formation location of the OC due to dynamical considerations \citep{Izidoro2015}. At their formation locations interior to the OC, small protoplanets can accrete material rapidly, before gap-opening by an OC, and subsequently migrate inward to their present-day locations via Type I. We therefore interpret our results as direct evidence for Type I migration of inner SEs and SNs. This scenario also provides natural explanations for 1) our observed invariance of the OC excess between SE and SN systems (c.f. Figure~\ref{fig:ratio_violin_main}) and 2) reports that inner SEs harbor WMFs of up to a few percent \citep[e.g.][Skinner et al. (under review)]{Rogers2025,Weisserman2026} as water vapor sublimated off of icy pebbles that drift inward from the outer disk can supply forming SEs without producing ``water-worlds'' with WMFs~$\gtrsim 10\%$ \citep{bitsch2021}.

\subsection{Stellar Mass Dependence of P(OC$|$ISP)/P(OC)}
In Section~\ref{sec:res_mass} we presented our measurements of P(OC$|$ISP)/P(OC) as a function of host star mass. We split our sample in bins of stellar mass separated at $0.6\, M_\odot$, corresponding to M and FGK dwarfs.
These stars exhibit unique occurrence rates of wide brown dwarf companions \citep{Grether2006,Sahlmann2011, Thompson2026_letter}, cold GGs \citep{Fulton2021,Mignon2025}, and ISPs \citep{Fressin2013, Dressing2015, Hsu2018, Hsu2020, Kunimoto2020, Gillis2026}, suggesting that the relative conditional occurrence of OCs given ISPs is similarly stellar mass-dependent. 

Many previous RV studies of the conditional occurrence rate of cold GGs given ISPs  were biased towards FGK dwarfs, with some M dwarf representation \citep[e.g.,][]{Bryan2016, Bryan2019, Bonomo2023, Bryan2024, Bonomo2025}. 
A consensus from these RV studies has emerged that when RV samples are debiased against over-representation of metal-poor stars, there exists a robust, positive correlation between cold GGs and ISPs around Sun-like stars \citep{Bryan2019,Zhu2024,Bryan2024}. Conversely, the same measurement around M dwarfs is less clear.
\citet{Bryan2019} reported  consistent occurrence rates for cold GGs around M dwarfs with ISPs compared to their full sample; namely, the excess of cold GGs in ISP systems is independent of host stellar mass. However, this result was based on an analysis of 11 M dwarfs out of a full sample of 59 stars. Follow up work in \citet{Bryan2025} measured cold GG occurrence rates around M dwarfs hosting ISPs with a larger sample of 85 M dwarfs. They calculated P(GG$|$ISP,MD) from RV data and compared this value to the field occurrence P(GG$|$MD) from \citet{Rosenthal_2021}, and found no correlation between ISPs and cold GGs around M dwarfs. This places tension on previous reports that cold GGs around M dwarfs are enhanced in ISP systems relative to the field \citep{Bryan2019}. 

By comparison, our results show that the relative occurrence of OCs around M dwarfs is greater than unity and appears slightly larger than around FGK dwarfs (i.e. by $1.41\pm0.78\times$; c.f. Figure~\ref{fig:ratio_violin_stellar_mass}). We report a positive correlation between ISP and OC occurrence around M dwarfs with OCs being enhanced by a factor of $2.44\pm 0.64$ in the presence of ISPs. This correlation appears stronger around M dwarfs than around FGK dwarfs, which still show an enhancement in OC occurrence given ISPs at the level of $1.73\pm 0.45$. Theoretical results predict that protoplanetary disks that are sufficiently massive in solids to form OCs can also form ISPs, a result that should hold regardless of host star mass \citep{Chachan_2023}. We therefore conclude that OCs around M and FGK dwarfs alike do not suppress the formation of ISPs, consistent with theoretical predictions from \cite{Chachan_2023}. 

However, FGK stars have more massive disks and an enhanced field occurrence of GGs \citep{Chachan_2023}. \citet{VanZandt2026} also found that OCs around FGK dwarfs are typically more massive than those around M dwarfs. As such, we attribute the apparent discrepancy between the relative occurrence rates shown in Figure~\ref{fig:ratio_violin_stellar_mass} as likely due to the difference in field OC occurrence between the stellar samples.

\subsection{Metallicity Dependence of P(OC$|$ISP)/P(OC)} \label{sec:dis_metallicity}
Stellar metallicity has been identified as an important factor affecting the correlation between ISPs and GGs \citep{Bryan2024, Zhu2024, Bryan2025, Bonomo2025}. GG formation is known to depend strongly on metallicity \citep{FischerValenti2005} and may also play a role in enabling small planet formation \citep[e.g.][]{Petigura2018, Turtelboom2026}. In particular, the positive correlation between ISPs and GGs is commonly argued to only be evident in metal-rich systems with [Fe/H] $\gtrsim 0$. 

We report an excess of OCs for the ALL ISP and SN planet samples in both our metal-rich and metal-poor stellar samples shown in Figure~\ref{fig:ratio_violin_metallicity}. This result for the metal-rich stellar sample is in line with previous RV surveys, with the ratio being larger than unity to 1.5$\sigma$. However, our finding of an enhanced relative OC occurrence in the presence of an ISP in metal-poor systems represents a divergence from the previous RV surveys. We highlight that we cannot directly compare our results with past RV surveys as our experiment is only sensitive to the relative OC occurrence and not absolute OC occurrence. 

Our observation of an enhanced relative OC occurrence rate around metal-poor stars may be partially attributed to the fact that our analysis is most sensitive to OCs more massive than $\sim 10\, M_{\rm Jup}$ (c.f. Figure~\ref{fig:sensitivity}). These objects are predominantly brown dwarfs (BDs) and, potentially, very-low-mass stars (VLMs). BDs and VLMs form by direct collapse of gas from the interstellar medium or outermost protoplanetary disk regions as in the gravitational instability (GI) paradigm of planet formation \citep{Boss1997, Stamatellos2009, Kratter2010}. GI, unlike core accretion which is likely responsible for producing the majority of GGs \citep{Pollack1996, Papaloizou1999}, is at most weakly dependent or independent of gas metallicity \citep{Carter2026}. This is reflected in the metallicity distributions of GG and BD host stars \citep{Giacalone2026}. As such, if our survey is preferentially probing BDs and VLMs in our inferred OC population rather than GGs, then we would not expect to see differences in the relative OC occurrence between our metal-rich and metal-poor samples. Additionally, given that the metallicity distribution of our sample is tightly clustered around [M/H]=0, the sample is heterogeneous and may be systematically biased due to insufficient calibrations, and individual [M/H] values have large uncertainties \citep[$\sim 0.3 \rm{dex}$,][]{recioblanco2023}, we acknowledge that our metal-poor and metal-rich stellar samples are likely cross-contaminated, reducing the statistical significance of any physical signal that may be present in the underlying populations.

\section{Summary \& Conclusions} \label{sec:conclusion}
This work presents the first analysis of the correlation between outer companions (OCs) with inner small planets (ISPs) using astrometry. Our work validates the excess of OCs in systems with ISPs--a consensus among previous radial velocity (RV) studies--an important milestone given the uniqueness of our dataset from Gaia DR2 and DR3, the $6.2\times$ increase in sample size compared to the largest RV surveys, and our homogeneous characterization of OC occurrence using a Bayesian hierarchical model. Our large sample size allows us to robustly study the correlation between OCs with subpopulations of super-Earths (SEs) and sub-Neptunes (SNs) and across stellar mass and metallicity. We summarize our main findings below.

\begin{itemize}
\item We find a statistically significant excess of OCs in systems hosting ISPs of P(OC$|$ISP)/P(OC)$=1.77\pm 0.49$, providing the first astrometric confirmation of this positive correlation.
\item We leverage our large sample size of 7,321 stars and establish that OC occurrence is similarly enhanced by a factor of $\sim 2$ in both inner SE-only and inner SN-only systems. We interpret this consistency as evidence that SEs and SNs alike formed rapidly and ex situ--though not necessarily beyond the snow line--before exterior OCs open gaps. Thus, we interpret our results as evidence for Type I migration of ISPs to their present-day orbits as described in Section~\ref{sec:formation}.
\item We find no significant difference in OC occurrence between SE and SN systems, suggesting that OCs do not preferentially favor the formation of any type of ISP and therefore are not strong drivers of ISP bulk compositions. 
\item The excess of OCs in systems with any type of ISP persists around both M and FGK dwarfs, with the excess of OCs appearing slightly stronger around M dwarfs by $1.41\pm0.78\times$ relative to FGK dwarfs.
\item Contrary to previous RV studies, we find no strong dependence of the OC--ISP correlation on stellar metallicity. We reconcile this result with past RV surveys by noting that this finding likely reflects that our astrometric analysis is most sensitive to substellar OCs (predominantly $\gtrsim 10\, M_{\rm Jup}$) whose formation is largely insensitive to metallicity. We also caution that the relative OC occurrence rates in our metal-rich versus metal-poor samples are not as robust as our other results due to poorly-matched stellar and astrometric parameters that emerge when slicing the samples, and because the metallicities used are from heterogenous sources that are tightly clustered around [M/H]=0 with large error bars of $\sim 0.3$ dex, thus likely diluting any potential underlying metallicity dependence.
\end{itemize}

It is important to note that the astrometric data used herein from Gaia DR2 and DR3 exhibits limited detection sensitivity to gas giant planets ($\lesssim 13\, M_{\rm Jup}$) at any orbital separation. To ensure that we are reporting robust inferences, we restrict our analyses to regions of the mass--semi-major axis parameter space in which our hierarchical model is largely insensitive to prior assumptions. Forthcoming epoch astrometry from Gaia DR4 will expand our results with anticipated higher detection sensitivity.
This study represents an important benchmark test for the utility of astrometry to tackle key questions in small planet formation theory and exoplanetary system architectures.

\begin{acknowledgments}
RC acknowledges the support of the Natural Sciences and Engineering Research Council of Canada (NSERC) and the Canada Research Chairs (CRC) Program [Award ID: CRC-2024-00372]. D.J.\ is supported by NRC Canada and by an NSERC Discovery Grant. 
J.S. acknowledges financial support from the Heising–Simons Foundation through the 51 Pegasi b Fellowship (grant \#2025-5885).

This work has made use of data from the European Space Agency (ESA) mission Gaia (\href{https://www.cosmos.esa.int/gaia}{https://www.cosmos.esa.int/gaia}), processed by the Gaia Data Processing and Analysis Consortium (DPAC),\\\href{https://www.cosmos.esa.int/web/gaia/dpac/consortium}{https://www.cosmos.esa.int/web/gaia/dpac/consortium.} This work made use of the Gaia Observation Forecast Tool (GOST). This research used the facilities of the Canadian Astronomy Data Centre operated by the National Research Council of Canada with the support of the Canadian Space Agency.
The authors acknowledge the use of OpenAI’s ChatGPT (GPT-6) model to assist with code debugging and figure formatting.

\end{acknowledgments}

\begin{contribution}

\end{contribution}

\section*{Data Availability}
The per-star posteriors can be browsed at \url{https://exopost.st} under the ``TOI" source option. 

\facilities{Gaia, ExoFOP}

\software{astropy \citep{2013A&A...558A..33A,2018AJ....156..123A,2022ApJ...935..167A},  scitkit-learn \citep{scikit-learn} SciPy \citep{2020SciPy-NMeth}, Octopodes.jl \citep{Thompson2026_letter}, Octofitter \citep{Thompson2023}, Pigeons.jl \citep{pigeons}, Julia \citep{julia_bezanson}}

\appendix

\section{Appendix I}\label{sec:cdf_appendix}
\begin{figure*}
    \centering
    \includegraphics[width=1\linewidth]{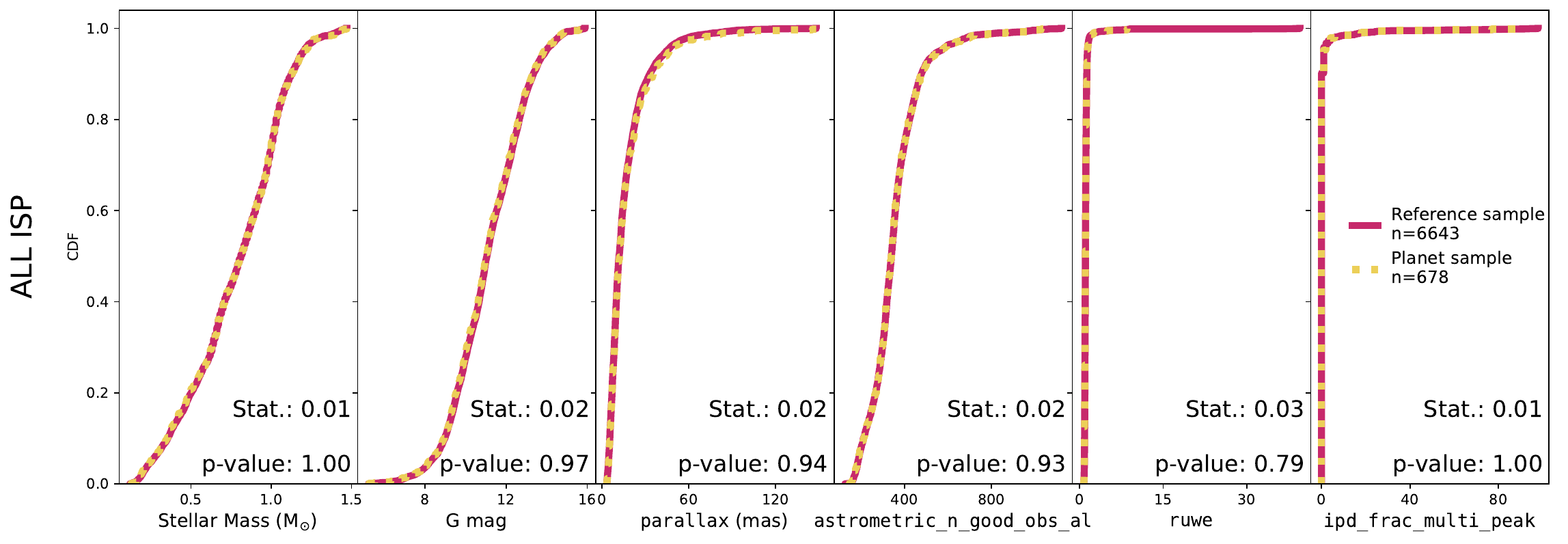}
\caption{Cumulative distribution of stellar and astrometric parameters of the planet samples (yellow) and their respective reference samples (pink) for the ALL ISP sample. This sample is not itself matched to its reference sample, but is the concatenation of the three individually matched samples SE, SN, SE+SN shown in Figure \ref{fig:cdfs}.}
    \label{fig:cdf_sesn}
\end{figure*}

In this appendix, we show additional CDFs for all subsamples resulting from cuts to our initial samples ALL ISP, SE, SN,
mentioned in the analysis, for completeness. 
With the exception of the SE+SN sample (Figure \ref{fig:cdf_sesn}), these samples are not themselves matched with the Nearest Neighbour method described in Section \ref{sec:ref-sample}, but are slices from the matched samples. 
We therefore expect some differences, but we present these CDFs to caveat our results, where the sensitivity could have some variation between the planet and reference sample. 
As in the CDFs presented in Section \ref{sec:ref-sample}, for each sample, we show the KS statistic and p-value, as well as the number of systems in the planet and its respective reference sample. 

\begin{figure*}
    \centering
    \includegraphics[width=1\linewidth]{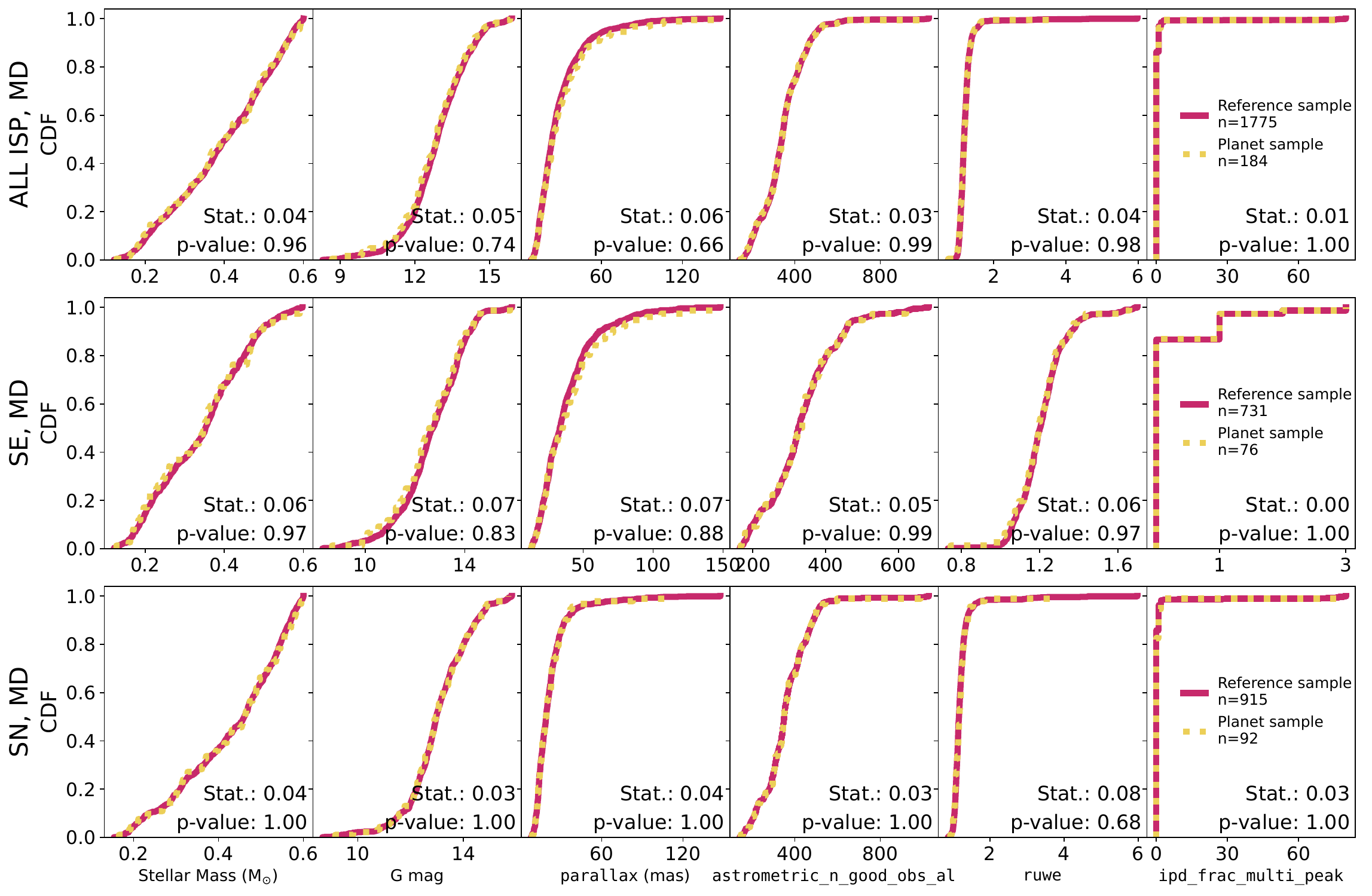}
    \caption{Resulting cumulative distribution of stellar and astrometric parameters of the planet samples (yellow) and their respective reference samples (pink) for the M dwarfs. These are the subsets of ALL ISP (top panel), SE (middle panel), and SN samples (bottom panel) stellar hosts that are M dwarfs.}
    \label{fig:cdf_mdwarfs}
\end{figure*}

\begin{figure*}
    \centering
    \includegraphics[width=1\linewidth]{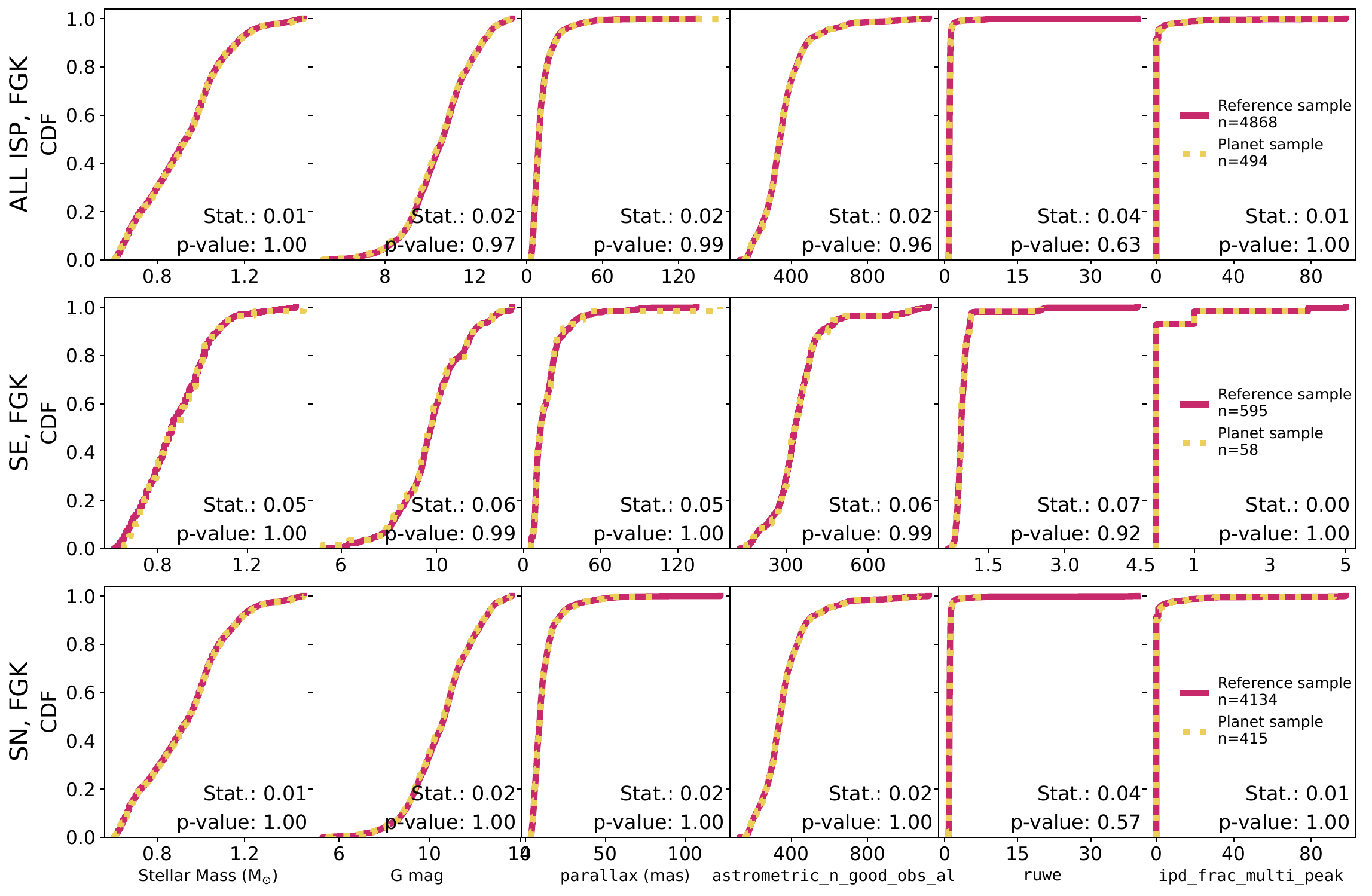}
    \caption{
    Resulting cumulative distribution of stellar and astrometric parameters of the planet samples (yellow) and their respective reference samples (pink) for the FGK dwarfs. These are the subsets of ALL ISP (top panel), SE (middle panel), and SN samples (bottom panel) stellar hosts that are FGK dwarfs.
    }
    \label{fig:cdf_fgk}
\end{figure*}

\begin{figure*}
    \centering
    \includegraphics[width=1\linewidth]{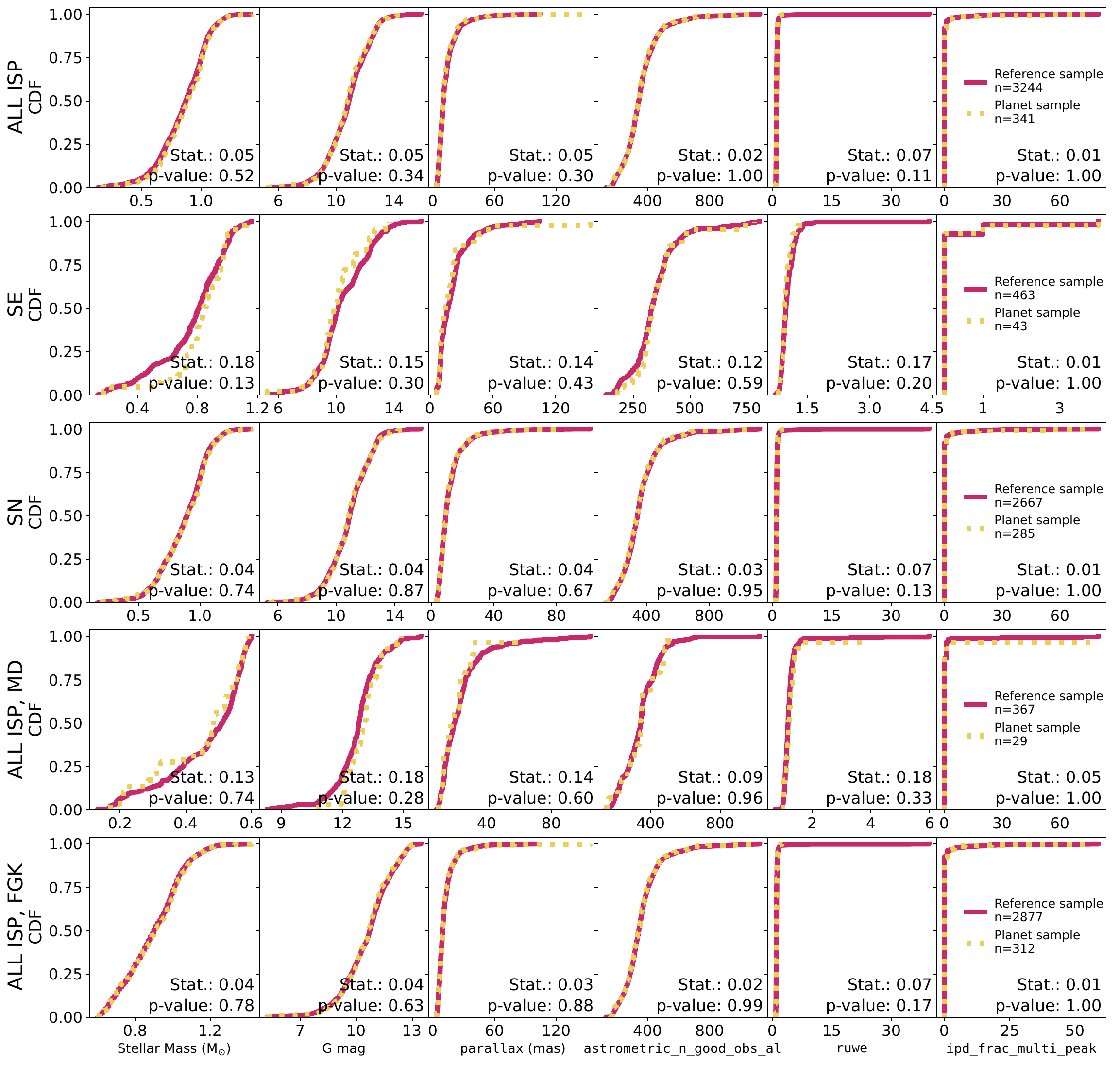}
    \caption{
        Resulting cumulative distribution of stellar and astrometric parameters of the planet samples (yellow) and their respective reference samples (pink) for the metal-rich hosts. These are the subsets of ALL ISP (first panel), SE (second panel), and SN samples (third panel) stellar hosts that are metal-rich. The bottom two panels show metal-rich M dwarfs and FGK dwarfs respectively.}
    \label{fig:cdf_MR}
\end{figure*}

\begin{figure*}
    \centering
    \includegraphics[width=1\linewidth]{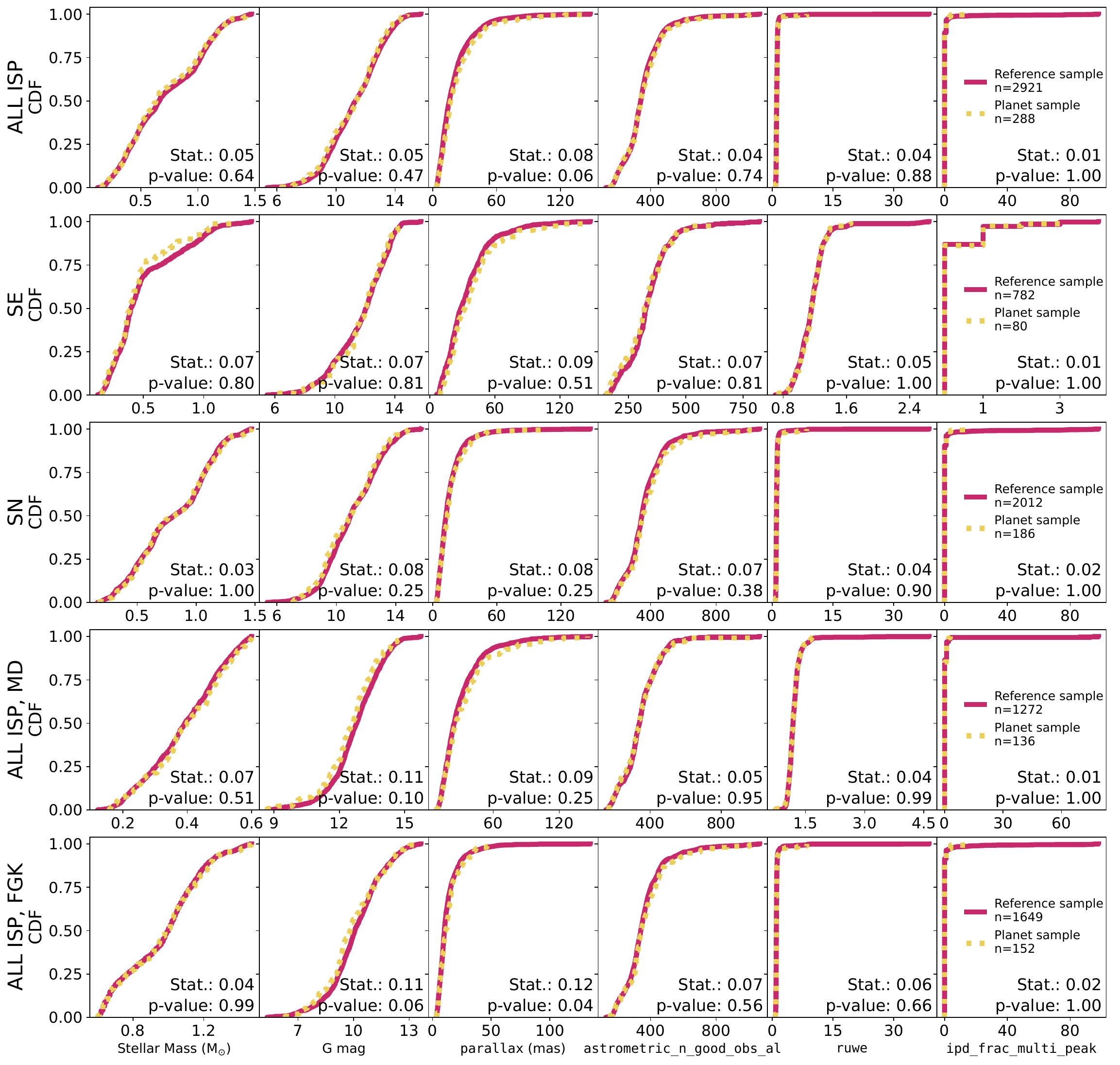}
    \caption{
        Resulting cumulative distribution of stellar and astrometric parameters of the planet samples (yellow) and their respective reference samples (pink) for the metal-poor hosts. These are the subsets of ALL ISP (first panel), SE (second panel), and SN samples (third panel) stellar hosts that are metal-poor. The bottom two panels show metal-poor M dwarfs and FGK dwarfs respectively.}
    \label{fig:cdf_MP}
\end{figure*}

In Figure \ref{fig:cdf_sesn}, we show the Nearest-Neighbour-matched SE+SN sample and the ALL ISP sample, which corresponds to the concatenation of the SE, SN and SE+SN samples. The samples that make up the ALL ISP samples are all individually matched. 
In Figure \ref{fig:cdf_mdwarfs} and \ref{fig:cdf_fgk}, we show the CDFs for our cuts in terms of stellar mass. We define M dwarfs as having $M_\star\leq0.6\, M_\odot$ and FGK stars as $M_\star>0.6\, M_\odot$.
In Figure \ref{fig:cdf_MR} and \ref{fig:cdf_MP}, we show the CDFs for the metallicity cuts. We obtain and calibrate metallicities from the Gaia mission, General Stellar Parametrizer-spectroscopy module 
\citep{recioblanco2023}, and additionally include metallicities calculated from empirical color-metallicity relations presented in \citet{Mann2013} and TIC metallicities
\citep{Stassun2019}.
This sample does not provide metallicity measurements for all the stars in our samples. 
Refer to Figure \ref{fig:mhdistribution} for the resulting hetereogenous metallicity distribution.
 
\bibliography{PASPsample701}{}
\bibliographystyle{aasjournalv7}

\end{document}